\documentclass[useAMS,usenatbib]{mn2e}
\usepackage{epsf}
\usepackage{amssymb}

\usepackage{color}
\definecolor{grey}{rgb}{0.4,0.6,0.6}
\definecolor{darkgreen}{rgb}{0.,0.7,0.}

\def\plotone#1{\centering \leavevmode
\epsfxsize=\columnwidth \epsfbox{#1}}
\def\plottwo#1#2{\centering \leavevmode
\epsfxsize=.99\columnwidth \epsfbox{#1} \hfil
\epsfxsize=.99\columnwidth \epsfbox{#2}}

\def\plottwos#1#2{\centering \leavevmode
\epsfxsize=.8\columnwidth \epsfbox{#1} \hfil
\epsfxsize=.8\columnwidth \epsfbox{#2}}
\def\plotones#1{\centering \leavevmode
\epsfxsize=.8\columnwidth \epsfbox{#1}}

\newcommand{\be}{\begin{equation}}
\newcommand{\ee}{\end{equation}}

\def\disp {\displaystyle}

\title[Elliptical galaxies: mass density slopes]{Mass density slope of elliptical galaxies from strong lensing and resolved stellar kinematics}
\author[Lyskova et al.]{N.~Lyskova,$^{1,2,3}$ 
E.~Churazov,$^{3,2}$ T. Naab$^{3}$ 
\newauthor \\
$^1$ National Research University Higher School of Economics, Myasnitskaya str. 20, Moscow 101000, Russia\\
$^2$ Space Research Institute (IKI), Profsoyuznaya 84/32, Moscow 117810, Russia\\ 
$^3$ Max-Planck-Institut f\"ur Astrophysik, Karl-Schwarzschild-Strasse 1, 85741
Garching, Germany\\
}

\begin{document}

\pagerange{\pageref{firstpage}--\pageref{lastpage}}
\pubyear{2017}

\maketitle

\label{firstpage}
\begin{abstract}

We discuss constraints on the mass density distribution (parameterized as $\rho\propto r^{-\gamma}$) in early-type galaxies provided by strong lensing and stellar kinematics data. The constraints come from mass measurements at two `pinch' radii. One `pinch' radius $r_1=2.2 R_{Einst}$ is defined such that the Einstein (i.e. aperture) mass can be converted to the spherical mass almost independently of the mass-model. Another `pinch' radius $r_2=R_{opt}$ is chosen so that the dynamical mass, derived from the line-of-sight velocity dispersion, is least sensitive to the anisotropy of stellar orbits. We verified the performance of this approach on a sample of simulated elliptical galaxies and on a sample of 15 SLACS lens galaxies at $0.01 \leq z \leq 0.35$, which  have already been analysed  in \cite{Barnabe.et.al.2011} by the self-consistent joint lensing and kinematic code.  For massive simulated galaxies the density slope $\gamma$ is recovered with an accuracy of $\sim 13$\%, unless $r_1$ and $r_2$ happen to be close to each other. For SLACS galaxies, we found good overall agreement with the results of  \citet{Barnabe.et.al.2011} with a sample-averaged slope $\gamma=2.1\pm0.05$.
 While the two-pinch-radii approach has larger statistical uncertainties, it is much simpler and uses only  few arithmetic operations with directly observable quantities.

\end{abstract}

\begin{keywords}
galaxies: elliptical and lenticular, cD - galaxies: kinematics and dynamics - gravitational lensing: strong
\end{keywords}

%

\sloppypar

\section{Introduction}

Over past two decades a number of dedicated systematic studies of galaxies acting as strong gravitational lenses has been initiated. For such systems a robust measurement of an aperture mass within  the Einstein radius  can be obtained. For ring-like and quadruply-imaged lenses the Einstein mass is  measured with especially high accuracy - up to a few percent - and this measurement is almost model-independent and assumption-free \citep{Kochanek.1991}. Unfortunately, strong gravitational lensing alone does not allow one to measure the slope of the mass profile of the lensing galaxy without additional assumptions due to the mass-sheet degeneracy \citep{Schneider.et.al.1992}.

The galaxy masses could be also independently probed via analysis of kinematics of gravitational potential tracers such as stars at small radii  \citep[e.g.,][]{Cappellari.et.al.2013, Cappellari.et.al.2015} and planetary nebulae and/or globular clusters at intermediate and large distances  \citep[e.g.,][]{Napolitano.et.al.2014, deLorenzi.et.al.2008,Morganti.et.al.2013}.
To construct a dynamical model for an early-type galaxy the available (projected) observables should be deprojected. Such analysis unavoidably invokes a number of assumptions on
(i)~the three-dimensional shape of a galaxy and often on
(ii)~the distribution of stellar (or other tracers') orbits.
Unfortunately, the anisotropy profile which characterizes the distribution of the orbits could not be unambiguously obtained from the most basic observables -  the surface brightness and  the projected velocity dispersion measurements. So far only complex and time-consuming methods 
to build distribution functions of early-type galaxies  using the Schwarzschild or the made-to-measure approaches  are able to constrain the anisotropy profile  \citep[e.g.,][]{Gebhardt.et.al.2003,Thomas.et.al.2004,Krajnovic.et.al.2005, van.den.Bosch.et.al.2008,Long.Mao.2012}.
Such  
analyses require high quality measurements of the light distribution and line-of-sight velocity moments (up to the fourth moment),  which are still often not enough to uniquely constrain all model parameters \citep[e.g.,][]{Dejonghe.Merritt.1992, Morganti.Gerhard.2012}. Without detailed kinematic profiles or for the analysis of large samples of galaxies, the Jeans equation is often in use with or without adopting the spherical symmetry \citep[e.g.,][]{Cappellari.et.al.2013}.   
To overcome the mass-anisotropy degeneracy one still needs to invoke additional assumptions (either on the anisotropy profile or on the mass profile) or to combine the method with other independent mass probes.

The joint analysis of lensing properties and line-of-sight kinematics  
has proven to be a powerful way of breaking both the mass-sheet degeneracy in lensing and the mass-anisotropy degeneracy in dynamics.
Here, we discuss  the possibility  to estimate the logarithmic gradient $\gamma$ of the mass density  combining strong lensing and kinematics data.
We concentrate  on the approaches, which are based on minimal observational data and use  
 only the direct observables, in particular, the surface brightness profile, the line-of-sight velocity dispersion profile, and the Einstein mass/radius.
Such methods can provide  quick and  robust estimates of masses and mass slopes of galaxies acting as strong gravitational lenses. They can be used,  for instance, for constraining the parameter space that needs to be explored to construct more detailed dynamical models. 

A widely used approach is based on a combination of an Einstein mass and a SDSS central velocity dispersion measurements with the assumption of an isotropic  velocity anisotropy profile. The SLACS team applied this methodology to a sample of $> 80$ nearby early-type galaxies  and found the mass density slope averaged over the sample to be close to isothermal \citep[see, e.g.,][]{Koopmans.et.al.2006,Auger.et.al.2010}. Another ready-to-use approach was presented in \cite{Agnello.et.al.2013} and is based on the joint use of the virial theorem and the lens equation. Using measurements of the Einstein radius, the surface brightness profile and the aperture velocity dispersion, the method allows one to obtain the logarithmic slope of the density with constrained dependence on orbital anisotropy.

Here, we present another easy-to-use approach that uses local values of the velocity dispersion profile instead of aperture measurements. The latter allows one to weaken the influence of the unknown anisotropy $\beta$ and to circumvent the mass-anisotropy degeneracy without apriori parametrization of $\beta(r)$.

This article is organized as follows. In Section~\ref{sec:method} we briefly describe procedures for recovering spherical masses from dynamics and strong lensing data. 
We validate the method on a set of simulated galaxies in Section~\ref{seq:sims}
and on a sample of real galaxies, already analysed with more sophisticated methods, in Section~\ref{sec:lensing_sample}.  
Section~\ref{seq:disc} summarizes our results.

\section{Description of the simple methods}
\label{sec:method}

We concentrate on two easy-to-use mass estimators designed for massive elliptical galaxies \citep{Churazov.et.al.2010, Lyskova.et.al.2015}. They are
based on minimal observational data, namely, on the surface brightness and projected velocity dispersion profiles and 
 are derived
from the spherical Jeans equation. The latter is valid for spherically
symmetric, dispersion-supported, collisionless, stationary
systems, in dynamical equilibrium, in the absence of
streaming motions.

Under the assumptions  of
(i) spherical symmetry,
(ii) approximately isothermal gravitational potential\footnote{Throughout this paper, `log' denotes natural logarithm} $\Phi(r)=V_c^2\log(r)+const$ (or in terms of masses, $M(r) \propto r$)
of the system and
(iii) constant velocity anisotropy $\beta$, one can manipulate the spherical Jeans equation in order to infer 
how the line-of-sight velocity dispersion profile $\sigma(R)$ of the stars (or other gravitational potential tracers) in the galaxy is expected to look like.
For a power law surface brightness $I(R) \propto R^{-2}$ one can show that the projected velocity dispersion
is independent of the radius and the distribution of stellar orbits \citep{Gerhard.1993}.
For a general surface brightness distribution the
relations between the circular speed and the local properties of $I(R)$ and $\sigma(R)$ for isotropic ($\beta = 0$), circular ($\beta = - \infty$) and radial ($\beta = 1$) stellar orbits are  \citep[see][]{Churazov.et.al.2010}:

\[
\sigma^{\rm iso}(R) = V_c(r) \frac{1}{\sqrt{1+\alpha+\zeta}} 
\]
\be
\sigma^{\rm circ}(R) = V_c(r) \sqrt{\frac{\alpha}{2 (1+\alpha+\zeta)}}
\label{eq:main} 
\ee
\[
\sigma^{\rm rad}(R) = V_c(r)\frac{1}{\sqrt{\left(\alpha+\zeta \right)^2+\delta-1}}, 
\]
where 
\be
\alpha\equiv-\frac{d\log I}{d\log R}, \ \ \zeta \equiv -\frac{d\log
  \sigma^2}{d\log R},\ \ \delta\equiv \frac{d^2\log[I\sigma^2]}{d
  (\log R)^2}.
\label{eq:agd}
\ee

It makes sense to choose a radius $R_{opt}$, optimal for mass estimation (sometimes called the `pinch' radius), at which the values of $\sigma^{\rm iso}(R)$, $\sigma^{\rm circ}(R)$ and $\sigma^{\rm rad}(R)$, derived for all three models spanning the range of possible anisotropies, are close to each other. At this radius
the local value of the line-of-sight velocity dispersion provides a measure of the circular speed $V_c(r) = \sqrt{GM(r)/r} $ or the dynamical mass $M(r)$ of the galaxy that is relatively insensitive to  the stellar  orbital distribution.
Ideally, one could define this radius as the point where all three curves given by eq.~(\ref{eq:main}) intersect. In reality the pairs of these three curves intersect at different radii, making the definition of the optimal radius ambiguous. For instance, one can use a provisional definition of the optimal radius $R_{\rm x}$, at which the root-mean-square deviation between three curves is minimal.

For general
spherical models  $R_{\rm x}$ is expected to lie not far from the radius $R_2$,
where the surface brightness declines as $R^{-2}$  (in other words, $\alpha = 2$).
In practice, as show our tests on analytical models and on simulated galaxies \citep[see, e.g.][]{Lyskova.et.al.2012,Lyskova.et.al.2015}, $R_2$ can be also used as the optimal radius for mass determination. It is easier to calculate than $R_{\rm x}$ as it is derived from the logarithmic slope of the surface brightness profile only.
So the circular speed can be expressed as
\be
V_c(R_2) \simeq \sqrt{(3+\zeta)}\sigma(R_2),
\label{eq:VcR2}
\ee
or in terms of masses
\be
M(R_2) \simeq \frac{(3+\zeta)\sigma^2(R_2)R_2}{G}.
\label{eq:MR2}
\ee

The above mass estimator is based on the {\it local} properties of the observed surface brightness distribution $I(R)$ and the velocity dispersion profile $\sigma(R)$ (rather than a single aperture\footnote{Throughout the paper, $\sigma(R_x)$ is the {\it local} value of the projected velocity dispersion at the radius $R_x$, and $ \sigma_{ap}(<R_x)  $ is the luminosity-weighted {\it aperture} velocity dispersion measured within $R_x$} velocity dispersion measurement $\sigma_{ap}$).
Let us note again, that eq.~(\ref{eq:MR2}) is obtained for spherical systems with the mass profile $M(r) \propto r$ and  constant anisotropy ($\beta(r) = const$).
Of course, real elliptical galaxies are never perfectly spherical, nor they have exactly $M(r) \propto r$ mass profile or constant anisotropy. Despite all these complications the above relation was shown to provide essentially unbiased masses of massive elliptical galaxies with the RMS scatter of $\simeq 12\%$ \citep[see tests in][]{Lyskova.et.al.2015}.

If  additionally we assume that the surface brightness distribution is described by the S\'{e}rsic profile $I(R) \propto \exp\left(-b_n (R/R_{\mathrm{eff}})^{1/n}\right)$, where $R_{\mathrm{eff}}$ is the effective (half-light) radius, then we find that
the velocity dispersions are close to each other at  $R_{\rm x} \simeq 0.5 - 0.6 R_{\mathrm{eff}}$.  For the de Vaucouleurs surface brightness profile (e.g., $n=4$) $R_{\rm x} \simeq 0.505 R_{\mathrm{eff}}$ and $\sigma(R_{opt}) \simeq 0.6V_c(R_{opt})$ (see Figure 3 in \citealt{Churazov.et.al.2010}), 
or in terms of masses,
\be
\displaystyle M(R_{\rm x})=\frac{1}{0.6^2}\frac{R_{\rm x}\sigma^2(R_{\rm x})}{G}.
\label{eq:Ropt}
\ee
  This mass estimator is primarily based on two observables - the effective radius and the {\it local}
value $\sigma(R_{\rm x})$
 of the velocity dispersion  at $R_{\rm x} \simeq 0.5 - 0.6 R_{\mathrm{eff}}$. The relation~(\ref{eq:Ropt}) was shown to recover masses at the optimal radius within 10\% accuracy for massive elliptical galaxies with  virial masses $10^{12}M_{\odot} \lesssim M_{vir} \lesssim 10^{14}M_{\odot}$ \citep{Chae.et.al.2012}.

The mass estimator from eq.~(\ref{eq:Ropt}) was derived for galaxies with the S\'{e}rsic surface brightness profile while the estimator from eq.~(\ref{eq:MR2}) is suitable for a general surface brightness distribution. Note, that determination of the effective radius (and the S\'{e}rsic index)
of real galaxies depends (sometimes strongly) on details of the analyses. $R_{\mathrm{eff}}$ could
vary depending on (i) whether it is measured
with or without extrapolation of surface brightness data, (ii) parametric form
of the stellar distribution profile used to fit the data, (iii)
radial range used to fit, for instance, the S\'{e}rsic profile, (iv)
quality of photometric data \citep[see, e.g.][]{Kormendy.et.al.2009,Cappellari.et.al.2013}.

\subsection{Mass density slope}
\label{seq:slope}

An analysis of the lensing effect allows one to measure the total mass of a lens enclosed in projection within the Einstein radius to a precision of a few percent (for symmetric quadruple-image lenses and for Einstein ring images) \citep[see ][]{Kochanek.1991, Bolton.et.al.2008}. 
The aperture mass
\be
\displaystyle M_{ap} = \int_0^{R_{ap}} 2 \pi R \int_{- \infty}^{+ \infty} \rho(R,z) dz dR
\ee
can be straightforwardly converted to spherical mass distribution under assumptions of spherical symmetry  and a power-law total density profile $\rho \propto r^{-\gamma}$ of a galaxy. The ratio between the Einstein mass (i.e., the projected mass within the Einstein radius $R_{Einst}$) to the spherical mass within $r = R_{Einst}$ is simply    
\be
\displaystyle \xi(\gamma) = \frac{M_{Einst}}{M(r=R_{Einst})} =  \frac{\sqrt{\pi}}{2}\frac{\Gamma\left[\frac{\gamma-1}{2}\right]}{\Gamma\left[\frac{\gamma}{2}\right]},
\ee
where $\displaystyle \Gamma[t] = \int_0^{\infty} x^{t-1}e^{-x}dx$ is the gamma-function.
Here it is assumed that the external convergence (i.e. a contribution of line-of-sight structures to a lensing signal) can be neglected. This is a viable assumption for low-redshift ($z \lesssim 1$) lenses \citep[see, e.g.][]{Wambsganss.et.al.2005, Treu.et.al.2009}. 
So the mass profile of a spherical system with $\rho \propto r^{-\gamma}$ can be written as
\be
M(r) = \frac{M_{Einst}}{\xi(\gamma)}\left( \frac{r}{R_{Einst}} \right)^{3-\gamma}.
\label{eq:Mgamma}
\ee
Or in terms of a circular speed
\be
\displaystyle V_c^2(r) = \frac{GM_{Einst}}{\xi(\gamma)R_{Einst}}\left( \frac{r}{R_{Einst}} \right)^{2-\gamma}.
\label{eq:Vcgamma}
\ee

\begin{figure*}
\plotone{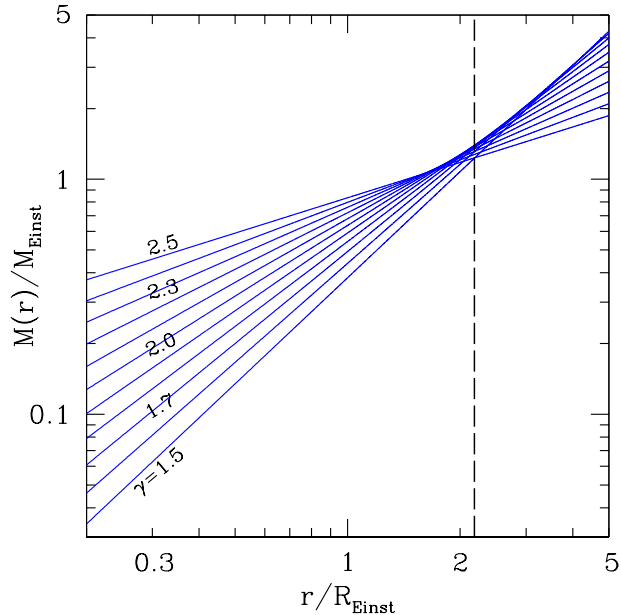}
\caption{Radial mass distribution $M(r)/M_{Einst}$ as a function of $r/R_{Einst}$ derived for power-law density profiles $\rho \propto r^{-\gamma}$ with $\gamma$ varying from $1.5$ to $2.5$ (eq.~(\ref{eq:Mgamma})). The dashed line marks the position of the `pinch' radius $2.18 R_{Einst}$ where mass deprojection is least sensitive to $\gamma$.   
\label{fig:mass_ratio}
}
\end{figure*}

An independent measurement of $M(r_0)$ at some radius $r_0$ in principle allows us to infer the total  density slope $\gamma$. As it can be seen from Figure~\ref{fig:mass_ratio}, measurements of 
$M(r_0)$ at radii $r_0 \lesssim R_{Einst}$ generally give  more stringent constraints on $\gamma$. For $r_0 \gtrsim 1.4-1.5 R_{Einst}$ the ratio $M(r_0)/M_{Einst}$ as a function of $\gamma$ become quite flat or non-monotonic, and in such a case  two mass estimates ($M(r_0)$ and $M_{Einst}$) are not enough to put stringent constraints on the total density slope. From Figure~\ref{fig:mass_ratio} it is also clear that for a physically motivated range of total density slopes $1.5 \le \gamma \le 2.5$ values of $M(r)$ for the given $M_{Einst}$ are rather
similar at about $2.0-2.2 R_{Einst}$.
This result suggests that somewhere near this radius one may recover the spherical mass of a system that is largely independent of the details of the total density distribution:
\be
 M(2.18 R_{Einst}) \approx \eta(\gamma) M_{Einst},
\label{eq:MfromEin}
\ee
   where $\eta$ varies from 1.23 to 1.4 when $1.5 \le \gamma \le 2.5$ (conservative estimate).
     If we assume that $\gamma$ is distributed as found in \cite{Auger.et.al.2010} for SLACS galaxies, namely, if we take the normal distribution with the mean $\left< \gamma \right> = 2.078$ and the scatter of $0.16$, then the mathematical expectation of $\disp M(2.18 R_{Einst})$ is $1.38M_{Einst}$ and the 68\% confidence interval ranges from $1.33M_{Einst}$ to $1.4M_{Einst}$. 


\subsection{Choosing the appropriate optimal radius}

For galaxies with surface brightness profiles described by the S\'{e}rsic law $I(R)=I(R_{\mathrm{eff}}) \exp\left(-b_n \left[ (R/R_{\mathrm{eff}})^{1/n}-1 \right] \right)$
the radius $R_2$ is related to $R_{\mathrm{eff}}$ via \citep{Graham.Driver.2005}:
\be
\displaystyle R_2 \simeq \left( \frac{2n}{b_n} \right)^n \simeq 1.2 R_{\mathrm{eff}}.
\label{eq:Sersic_R2}
\ee  
In principle, both $R_2 \simeq 1.2 R_{\mathrm{eff}}$ and $R_{\rm x} \simeq (0.5-0.6) R_{\mathrm{eff}} $ can be used for the mass estimation. Moreover, 
  there exist  a range of radii at which the resulting mass estimate is not very sensitive to the anisotropy profile. This is especially true for galaxies with relatively large S\'{e}rsic indices. For example, if $I(R)$ is approximately $\propto R^{-2}$  over some range of radii, then any radius in this range can  equally well serve as the optimal one for mass estimation. For galaxies with $n \lesssim 8-10 $ the curves in eq.~(\ref{eq:main}), used to determine the optimal radius, form a `triangle'. In principle, any radius within this `triangle' can be used for estimating the mass  with comparable scatter, though the bias could be different \citep[see, for example, Figure 3 in][where  shown the comparison between performances of mass estimators at $R_2$ and at $\simeq (0.5-0.6) R_{\mathrm{eff}} $ (called $R_{sweet}$ there)]{Lyskova.et.al.2015}. 
  
  This observation gives one a freedom in choosing the `right' pair of radii ($R_{Einst}$ and $R_{opt}$; $R_{Einst}$ is in practice fixed) for the density  slope estimation.  If, for example, the Einstein radius $R_{Einst} \sim 0.5 R_{\mathrm{eff}}$, then $2.2 R_{Einst} \sim 1.1 R_{\mathrm{eff}}$ and $R_2$ happen to be close to each other, what makes the slope estimation uncertain. On the other hand, $R_{opt} \simeq 0.5-0.6 R_{\mathrm{eff}}$ and $2.2 R_{Einst} \sim 1.1 R_{\mathrm{eff}}$ allow one to determine the average slope of the mass profile more accurately.

\section{Masses and mass density slopes of simulated galaxies}
\label{seq:sims}

Modern numerical simulations  of  galaxy  formation  are  now  able to produce realistic galaxy populations and to reproduce observed galaxy properties quite well \citep[see, e.g.][]{Naab.Ostriker.2017} . It makes model galaxies extremely valuable for testing 
different approaches designed to recover intrinsic physical properties of galaxies from limited observational information, including methods for mass determination. 
We make use of the 
simulations of the entire formation history of galaxies in a full cosmological context, described in detail in \cite{Oser.et.al.2010, Oser.et.al.2012}, to access  validity and performance of mass estimators discussed in Section~\ref{sec:method}. 
Here we briefly list  properties of model galaxies relevant for our studies.
 The simulations were carried out with the following cosmological parameters (in standard notation): $h = 0.72$, $\Omega_b$ = 0.044, $\Omega_{DM} = 0.216$, $\Omega_{\Lambda}$,  $\sigma_8 =0.77$ and initial slope of power spectrum $n_s = 0.95$ (WMAP3 cosmology, \citealt{Spergel.et.al.2007}).

The present day virial halo masses of the simulated galaxies range from
$2\times 10^{11} \mathrm{M}_{\odot}$ to $4\times 10^{13} \mathrm{M}_{\odot}$  within $R_{vir} = R_{200}$.
The central galaxy stellar masses vary from   $2\times 10^{10} \mathrm{M}_{\odot}$ to  $6\times 10^{11} \mathrm{M}_{\odot}$. Typical effective radii (projected half-light = half-mass radii) are $R_{\rm eff} \simeq 1-7$ kpc \citep{Oser.et.al.2010,Naab.et.al.2014}. The co-moving softening lengths for stellar particles in the resimulations is $\simeq 0.5$ kpc.
All galaxies are well resolved with $\simeq 1.4\times 10^4 - 2\times 10^6$ particles within the virial radius.
 The surface brightness of the model galaxies is generally well represented by a S\'ersic profile with a `core'  \citep{Wu.et.al.2014}, and most of the objects have large values of S\'{e}rsic indices ($n > 10$ or even $n \gg 10$), so that these galaxies have more power-law-like surface brightness profiles than real elliptical galaxies \cite[see, e.g.][]{Krajnovic.et.al.2013}.
Total density profiles of the simulated galaxies are mostly reasonably well described by a single power law with an super-isothermal average density slope $\left< \gamma \right> \simeq 2.3 \pm 0.28 $  \citep{Remus.et.al.2017}.

It has been demonstrated in \cite{Oser.et.al.2012} that the massive model galaxies have structural properties very similar to observed early-type galaxies, i.e. they follow the observed scaling relations and their evolution with redshift. Namely, the velocity, velocity dispersion, and higher velocity moments of the simulated galaxies show a diversity similar to observed kinematic
maps of early-type galaxies in the ATLAS$^{\rm 3D}$ survey \citep{Naab.et.al.2014}, which in turn  provides the most complete panoramic view on the
properties of 260 local early-type galaxies in a volume limited
sample covering different environments within a distance
of $\sim 42$ Mpc \citep{Cappellari.et.al.2011}.

We have considered each model galaxy in 3 independent projections making the final sample 3 times larger (195 objects in total).  
\subsection{Mass estimate from the Einstein mass measurement}

\begin{figure*}
\plotone{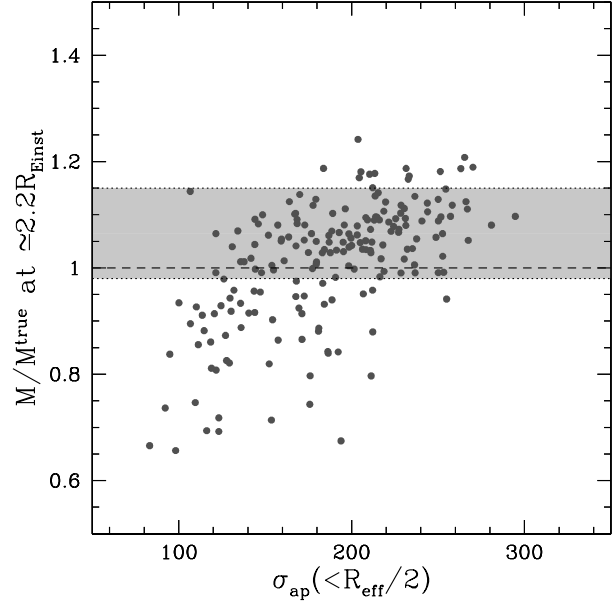}
\caption{The ratio between the estimated  spherical mass within  $\simeq 2.2 R_{Einst}$  (eq.~(\ref{eq:MfromEin})) and the true  value for simulated galaxies. The ratio is plotted as a function of the central velocity dispersion (aperture dispersion measured within $1/2$ effective radius). Here we consider the full sample of simulated galaxies in 3 independent projections (195 objects in total) including merging objects.  The grey area indicates the expected spread in  mass estimates  at $\simeq 2.2 R_{Einst}$  for $1.5 \le \gamma \le 2.5$.  The dashed line shows the expected value of $M(2.2R_{Einst}) \simeq 1.38 M_{Einst}$ for $\left< \gamma \right> = 2.078$, which corresponds to the average slope of the mass density profile for SLACS galaxies (\citealt{Auger.et.al.2010}). 
\label{fig:lens_sweet_spot}
}
\end{figure*}

First we test an accuracy and a scatter of  the `lensing' mass estimator at $\simeq 2.2 R_{Einst}$ (eq.~(\ref{eq:MfromEin})) which relies only on the Einstein mass measurement. We `mimic' lensing observations simply by assigning an Einstein radius to each object, where $R_{Einst}$ are pseudo-randomly chosen to roughly match the observed distribution of the ratio $R_{Einst}/R_{\mathrm{eff}}$ for nearby galaxies \citep{Auger.et.al.2009}, i.e. the $R_{Einst}$-distribution is peaked around $ 0.5 - 0.6 R_{\mathrm{eff}}$.
Then we calculate a projected mass within the Einstein radius $M_{Einst}$ and estimate a spherical mass at $\simeq 2.2 R_{Einst}$ for a given galaxy (and a given orientation)  using eq.~(\ref{eq:MfromEin}). The performance of this estimator is shown in Figure \ref{fig:lens_sweet_spot}.  The expected uncertainty in mass estimation is marked  as the grey area. The estimator works well for majority of  massive galaxies with the central velocity dispersions $\sigma_{ap}(<R_{\mathrm{eff}}/2)  \gtrsim 200 $ km s$^{-1}$, measured within one-half of the effective radius.
Substantial underestimation of mass is found for simulated objects with  the pinch radius $r_1 \simeq 2.2 R_{Einst}$ lying in the vicinity of the radius where the circular velocity curve reaches its maximum and starts to decline.

\subsection{Mass slope from joint lensing and dynamics analysis}

\begin{figure*}
\plotone{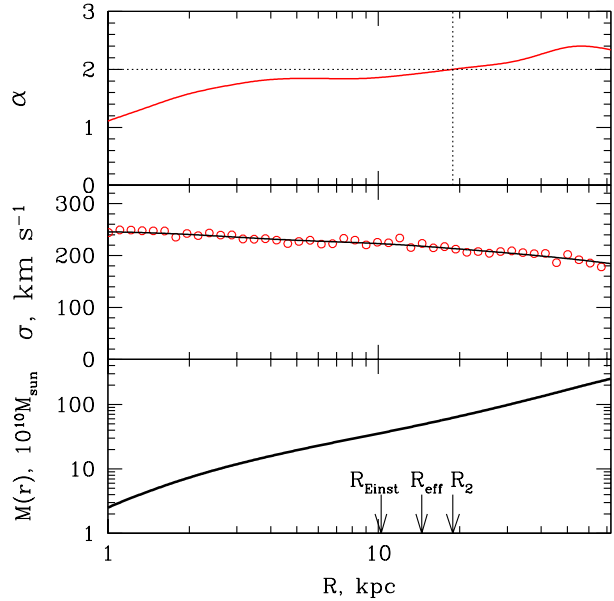}
\caption{ The surface brightness slope $\alpha  = - d\log I/d\log R$, the line-of-sight velocity dispersion profile $\sigma(R)$ and the  total mass $M(r)$ of a typical massive simulated galaxy (with stellar mass $M_{stars} = 3.8 \times 10^{11}M_{\odot}$ and virial mass $M_{vir} = 1.3 \times 10^{13}M_{\odot}$). Arrows on the lower panel mark the Einstein $R_{Einst}$ and effective $R_{\mathrm{eff}}$ radii, and $R_2$ at which $\alpha = 2$. 
\label{fig:typical_profile}
}
\end{figure*}

Now we estimate the density slope $\gamma$ from a combination of eqs.~(\ref{eq:Mgamma}) and~(\ref{eq:MR2}). We do this exercise for the subsample of  massive simulated galaxies with $\sigma(R_{\mathrm{eff}})>150$ km s${^{-1}}$ where merging and oblate objects seen along the rotation axis are excluded. For this subsample (106 objects in total) the simple $R_2$-estimator (eq.~(\ref{eq:MR2})) shows the best performance allowing to recover an almost unbiased estimate of the  mass (within $R_2$) with rms-scatter of $\simeq 10-12\%$ \citep[for more details, see][]{Lyskova.et.al.2012, Lyskova.et.al.2015}.
For simulated galaxies under consideration, $R_2$ is typically $\sim 1-2$ effective radii, and $R_{Einst} \sim 0.5-0.6\, R_{\mathrm{eff}}$. Figure~\ref{fig:typical_profile} illustrates the surface brightness slope $\alpha $ (see eq.~(\ref{eq:agd})), the line-of-sight velocity dispersion profile and the  total mass of a typical massive simulated galaxy (with stellar mass $M_{stars} = 3.8 \times 10^{11}M_{\odot}$ and virial mass $M_{vir} = 1.3 \times 10^{13}M_{\odot}$). Arrows on the lower panel mark the Einstein and effective radii, and $R_2$.    For each object  we determine $\gamma$ under assumptions of spherical symmetry and a power-law density profile from
\be
\displaystyle  \frac{(3+\zeta)\sigma^2(R_2)R_2}{G} \simeq M(R_2) = \frac{M_{Einst}}{\xi(\gamma)}\left( \frac{R_2}{R_{Einst}} \right)^{3-\gamma}
\label{eq:obvious}
\ee
if this equation has a unique solution in the range $1.5 \leq \gamma \leq 2.5$ or, in other words, if $R_2$ doesn't lie close to $2.2 R_{Einst}$. Namely, we exclude measurements with $0.7 \le R_2/(2.2R_{Einst}) \le 1.3$ (or $1.5 \le R_2/R_{Einst} \le 2.8$) (see Figure ~\ref{fig:mass_ratio}).

\begin{figure*}
\plottwo{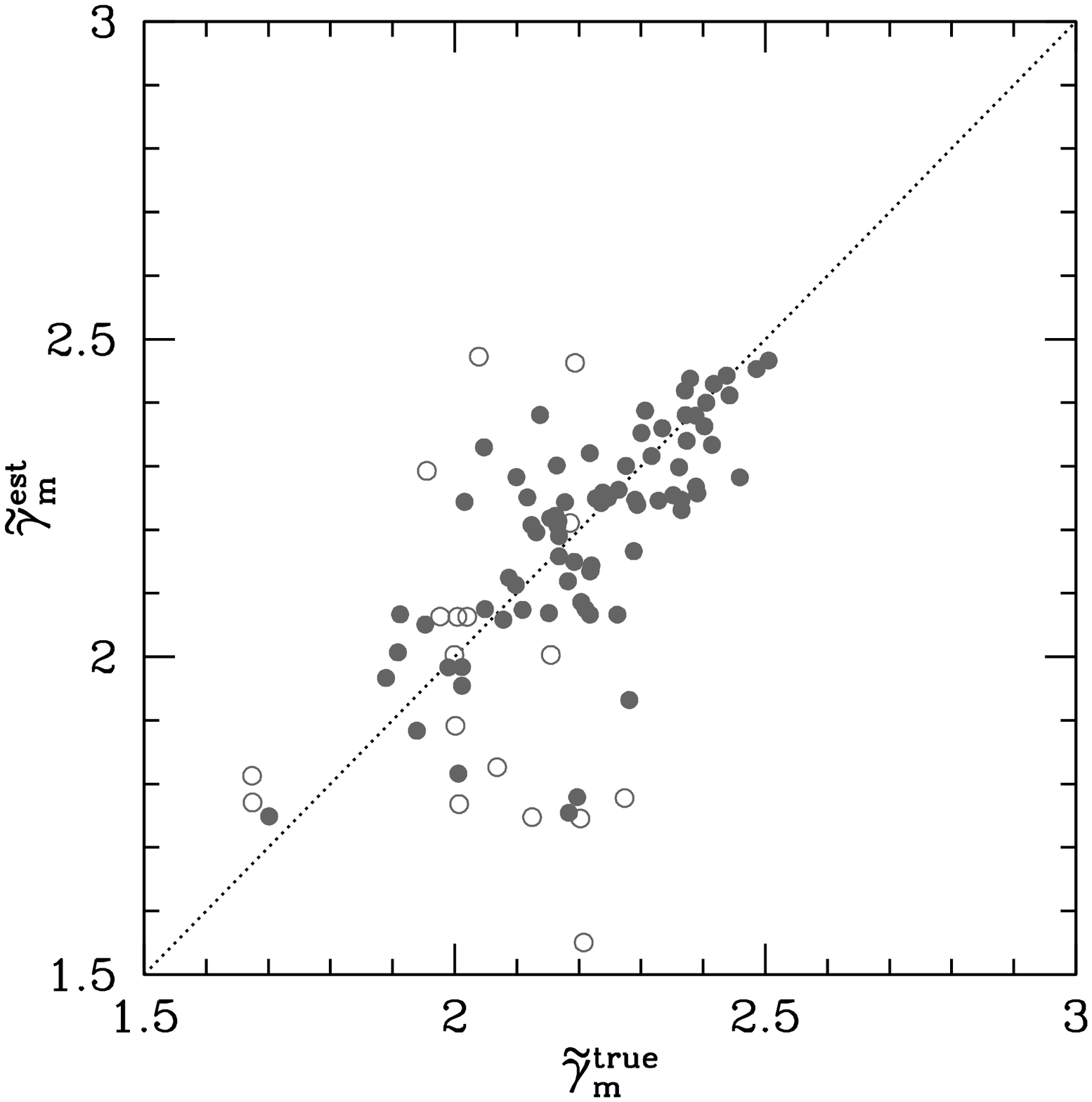}{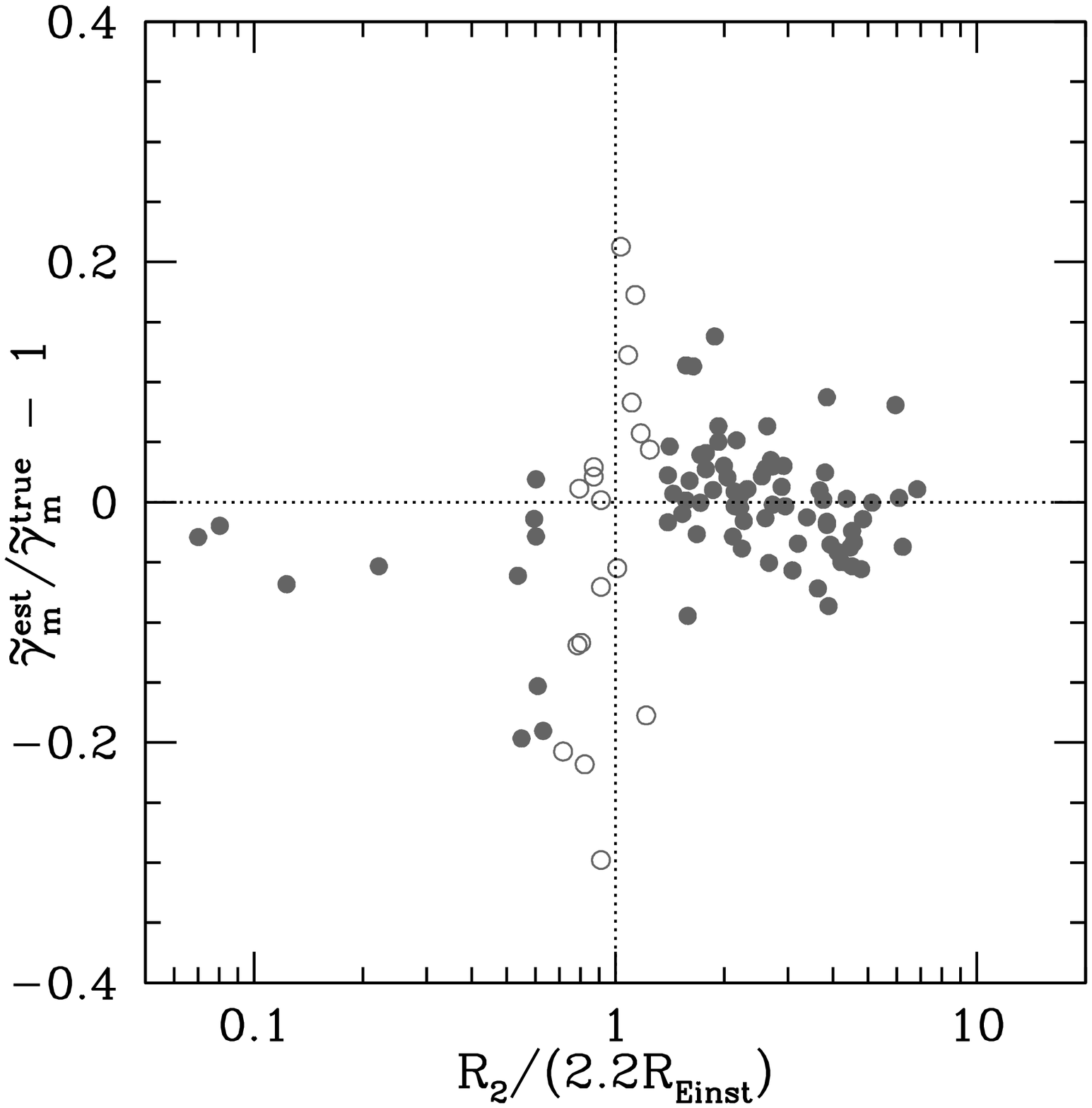}
\caption{Left: Density slopes for simulated galaxies estimated via eq.~(\ref{eq:obvious}) in comparison with the true average mass-weighted density slopes $\tilde{\gamma}_m = 3 - \log \left(M_2/ M_1\right)/ \log \left(r_2/r_1\right)$, where $M_2 = M(2.2R_{Einst})$, $r_2 = 2.2R_{Einst}$, $M_1 = M(R_2)$, $r_1 = R_2$. Open symbols denote (excluded from the analysis) objects with $0.7 \le R_2/(2.2R_{Einst}) \le 1.3$ for which the largest deviations are observed. Right: Influence of $R_2/(2.2R_{Einst})$ on the accuracy of recovered density slopes.     
\label{fig:sim_mass_slope}
}
\end{figure*}

Since both the gravitational lensing and stellar dynamics measure the total mass (rather than density) enclosed within different radii, we adopt here the following definition of the  mass-weighted density slope \citep{Dutton.Treu.2014}:
\be
\displaystyle \gamma_m(r) = -\frac{1}{M(r)} \int_0^r \gamma(x) 4\pi x^2 \rho(x) d x = 3-\frac{4\pi r^3 \rho(r)}{M(r)}.
\label{eq:gamma_m}
\ee
The mass-weighted slope is directly related to the local logarithmic slopes of the total mass $M(r)$ and circular speed $V_c(r)$ profiles:
\be
\displaystyle \gamma_m(r) = 3 - \frac{d \log M}{d \log r} = 2 - 2\frac{d \log V_c}{d \log r}. 
\label{eq:gamma_m2}
\ee
Note, that $\gamma \equiv \gamma_m$ only in the case of a power-law density distribution $\rho \propto r^{-\gamma}.$

By design the discussed method recovers the slope between two mass estimates: the `dynamical' one $M(R_2)$ and 
 the `lensing' one $M(2.2R_{Einst})$. So we introduce the average mass-weighted density slope  as
\be
\displaystyle \tilde{\gamma}_m = 3 - \log \left(\frac{M(2.2R_{Einst})}{ M(R_2)}\right)/ \log \left(\frac{2.2R_{Einst}}{R_2}\right).
\label{eq:mass_slope}
\ee

Figure~\ref{fig:sim_mass_slope} shows the resulting $\gamma$-estimates (filled circles) in comparison with the true average mass-weighted density slope (eq.~(\ref{eq:mass_slope})).  
Excluded from the analysis objects with $0.7 \le R_2/(2.2R_{Einst}) \le 1.3$ are shown as open circles.
As seen from Figure~\ref{fig:sim_mass_slope}, the general agreement between our $\gamma$-estimates  and the true values is good, and the rms-scatter is $\simeq 13$ per cent. The largest deviations are observed for objects with $R_2 \sim 1.3 R_{Einst}$ since this radius lies close to the point of intersection of mass profiles for $\gamma > 2$ (see Figure~\ref{fig:mass_ratio}), and the simulated galaxies under consideration are mostly super-isothermal as mentioned above.  So small deviations in $M(R_2)$ result in noticeable change in $\gamma$ derived from eq.~(\ref{eq:obvious}).  On general, such the approach gives a reasonable estimate of the  average mass-weighted density slope between $R_2$ and $2.2 R_{Einst}$. The main uncertainty-driver is the closeness of $R_2$ and $2.2 R_{Einst}$, as clearly seen from Figure~\ref{fig:sim_mass_slope} (right panel). If these two radii are  close to each other then  $\gamma$ resulting from eq.~(\ref{eq:obvious}) is not unique or has large error bars. On the other hand, if $R_2$ happens to lie sufficiently close to  $2.2 R_{Einst}$ one has two independent mass probes at the same radius.  Comparison between these mass estimates might in principle be used for more accurate assessment of uncertainties/possible biases. Although for the subsample of most massive ($ \sigma_{ap}(< R_{\mathrm{eff}}/2)  \gtrsim 200 $ km $s^{-1}$)  simulated galaxies under consideration the  mass estimator based on the lensing data  has a smaller scatter than the dynamical one.

 Although for the joint lensing and dynamics analysis presented in this Section, the `lensing' mass estimator $M(2.2R_{Einst})$ is not used explicitly,
it helps in interpreting the meaning of the inferred $\gamma$. Based on $\sigma(R_2)$, $R_{Einst}$ and $M_{Einst}$ measurements  the proposed above method recovers the effective slope of the mass profile $\tilde{\gamma}_m$ between $R_2$ and $2.2 R_{Einst}$ (eq.~(\ref{eq:mass_slope})).

 As mentioned above, the light distribution of simulated galaxies used for our tests is more power-law-like compared to real objects.  For this reason we estimate the total density slope using only $R_2$ as the optimal radius for mass evaluation  and avoiding determination of the effective radius from the S\'{e}rsic profile fitting.

\subsection{Comparison with other simple approaches}

We have also compared our approach for the density exponent estimation with the method discussed in  \cite{Koopmans.et.al.2006} which is widely used for  joint  strong  lensing  and  dynamics  analyses of surveys. The logarithmic density slope (inside $\approx$ the Einstein radius) is  derived from the Einstein mass measurement, the SDSS aperture stellar velocity dispersion, and the
surface brightness distribution. 
We apply the Koopmans et al. methodology for the same subsample of 106 relatively massive simulated galaxies with $\sigma(R_{\rm eff})>150$ km s${^{-1}}$. Merging and oblate systems seen along the rotation axis were excluded. 
Namely, we first fit the surface brightness distribution with a S\'{e}rsic profile in the radial range from the softening length ($ \simeq 0.5$ kpc ) up to 5 effective radii. 
We then solve the spherical Jeans equation assuming isotropic distribution of stellar orbits and power-law total density distribution, 
and for each logarithmic density
slope $1<\gamma<3$ derive the projected velocity dispersion profile as a function of radius. Finally, we calculate the aperture velocity dispersion within $R_2$, compare it with the `real' (extracted from simulations) value, find minimum discrepancy between these two aperture dispersions and the corresponding total density slope.

\begin{figure*}
\plottwo{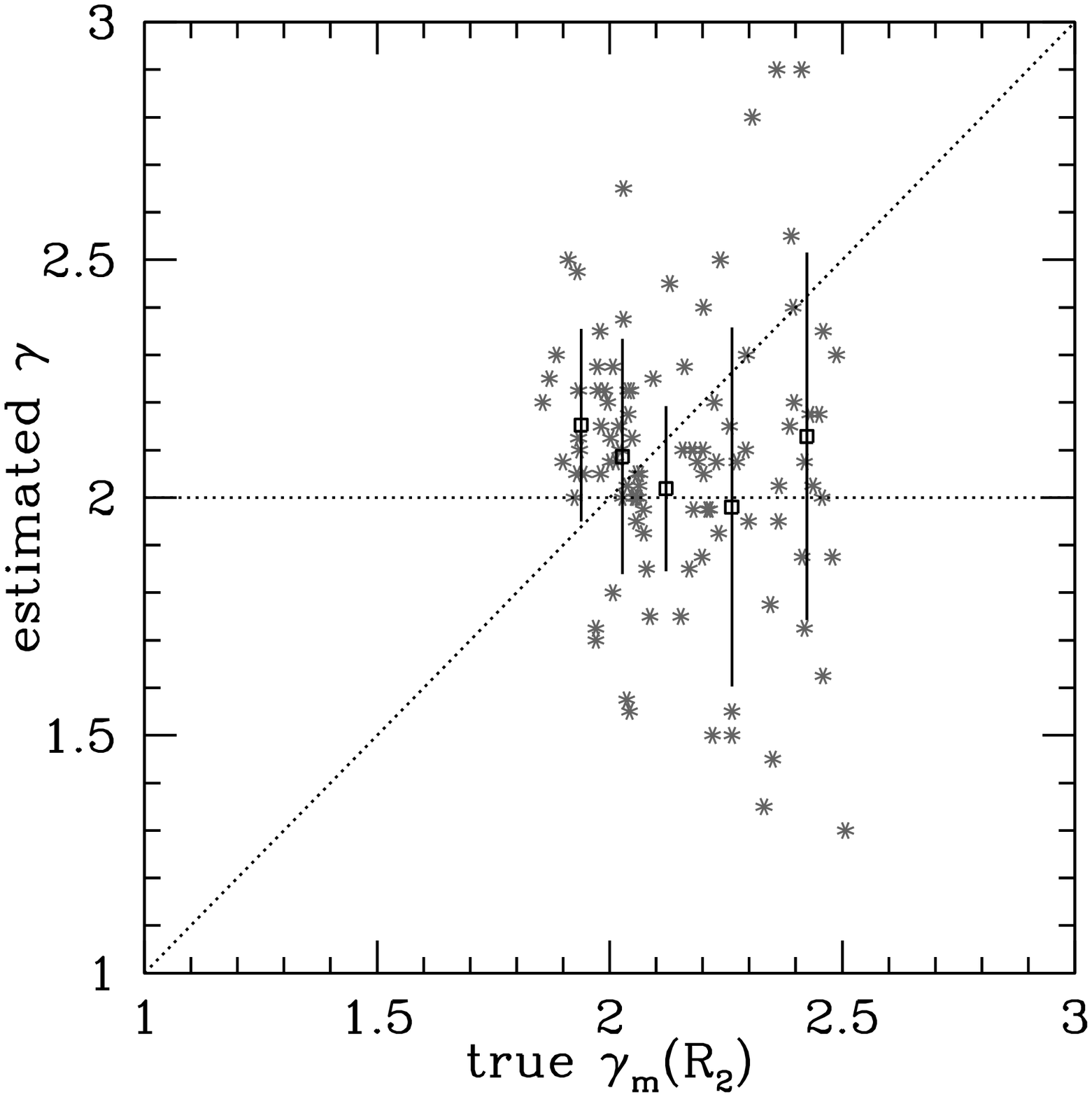}{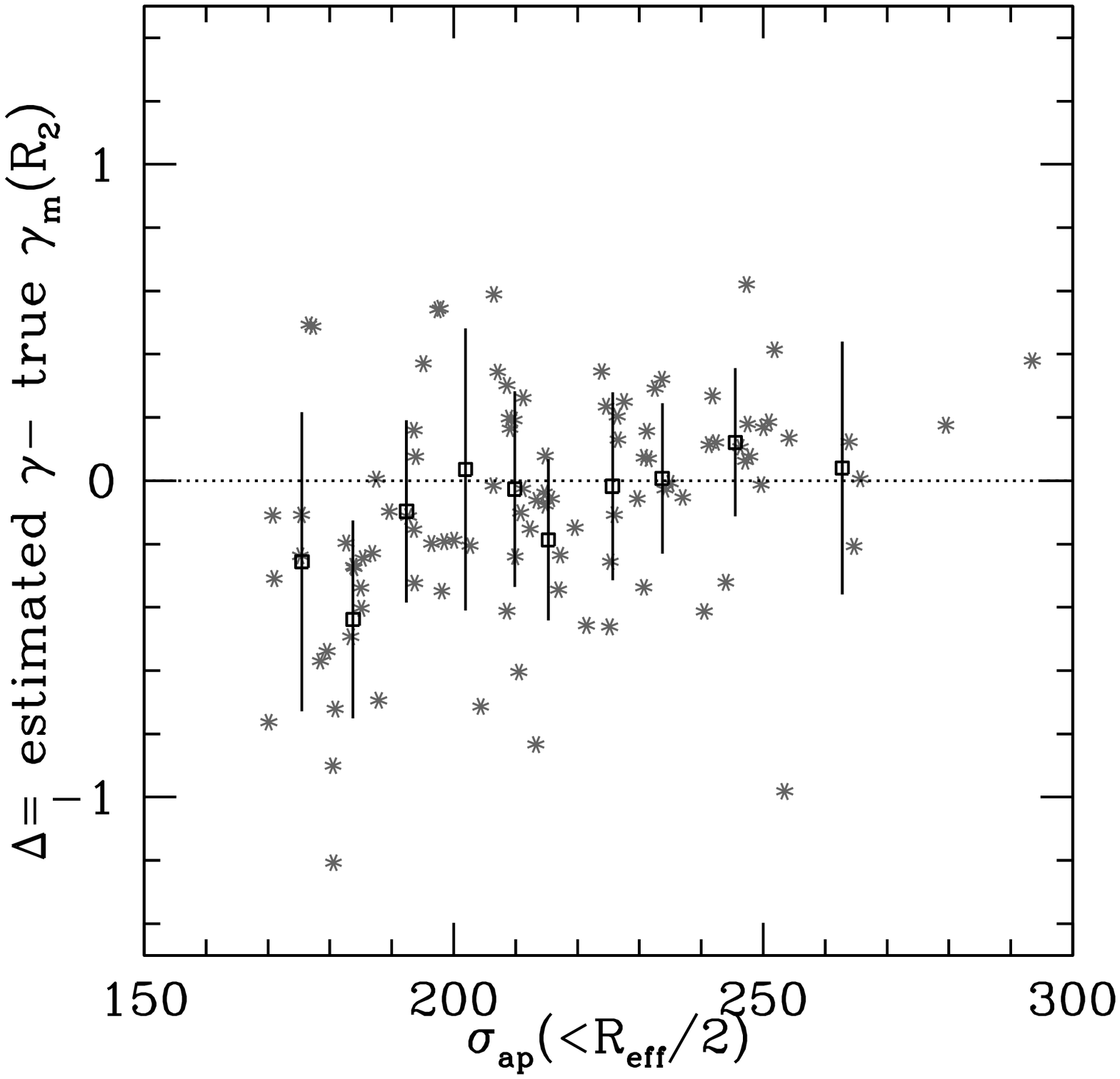}
\caption{Left: Density slopes recovered from the aperture velocity dispersion within $R_2$ (following the \citealt{Koopmans.et.al.2006} approach) as a function of the true mass-weighted density slopes $\gamma_m$ measured at $R_2$. Grey star symbols ($*$) are individual measurements. Black open squares with error bars show the arithmetic average of estimated density slopes binned by the value of true $\gamma_m(R_2)$. Right: Density slopes, estimated from the aperture velocity dispersion within $R_2$ (typically $R_2 \sim R_{\mathrm{eff}}$), as a function of the central (measured within $R_{\mathrm{eff}}/2$) stellar velocity dispersion.   Black open squares with error bars show the arithmetic average of estimated density slopes binned by the velocity dispersion values.  
\label{fig:sim_mass_slope_Koopmans}
}
\end{figure*}

The result is presented in Figure~\ref{fig:sim_mass_slope_Koopmans} where the left panel shows the comparison of the estimated density slope with the true mass-weighted density slope (eq.~(\ref{eq:gamma_m})-(\ref{eq:gamma_m2}))  which is expected to be a good proxy for the density slopes derived from the Koopmans et al. analysis \citep{Dutton.Treu.2014, Sonnenfeld.et.al.2013}.

Figure~\ref{fig:sim_mass_slope_Koopmans} shows almost no correlation between the density exponent estimates and the true (mass-weighted) density slopes for individual measurements. Note that the average estimated density slope is always close to $2-2.1$, and for most massive objects in the sample with central velocity dispersion $\gtrsim 200 $ km s$^{-1}$ the deviation of the estimated $\gamma_m$ from the true value is  on average close to zero (Figure~\ref{fig:sim_mass_slope_Koopmans}, right panel, black points with errorbars show the average deviation and corresponding RMS-scatter as a function of the central velocity dispersion) though the scatter is still quite large.  Note, that the spread in estimated density slopes (the estimated $\gamma$ varies from 1.3 to 2.9) is somewhat larger than found for SLACS lens galaxies \citep[][]{Auger.et.al.2010}, where the density slopes for individual lenses vary from 1.6 to 2.7. This might be caused by the broader mass range in the simulations compared to the one probed by the
SLACS sample. For simulated objects with the central velocity dispersion $\gtrsim 200 $ km s$^{-1}$ the spread becomes consistent with observations.

As the SDSS apertures for nearby galaxies rarely extend up to $R_2 \sim R_{\mathrm{eff}}$ but rather a some fraction of it, we also tested Koopmans et al. approach on smaller apertures. Main results are the same as discussed above: the average estimated density exponent is always close to $2-2.1$ and the scatter between the true and estimated density slopes preserves to be significant. \cite{Xu.et.al.2017} tested the Koopmans et al. approach on a larger sample of model galaxies in the Illustris simulations and reached qualitatively the same conclusion (see their Figure 17): a moderate disagreement exists between the simple estimate of the density slope and the intrinsic one especially for galaxies with $\sigma_{ap}(<R_{\mathrm{eff}}/2) \lesssim 200-250$ km s$^{-1}$. Also the Koopmans et al. estimator shows smaller scatter when the intrinsic mass-weighted density slope is close to $2$.

The main sources of systematic biases in the Koopmans et al. approach are (i) the assumption of isotropy which affects considerably the aperture velocity dispersions calculated form the spherical Jeans equation, and (ii) the power-law total density profile at $R \lesssim R_{\mathrm{eff}}$. The aperture dispersion seems to be more sensitive to the details of the anisotropy profile than the
dispersion at a single `optimal' radius where sensitivity to the unknown anisotropy is minimal, unless the aperture is really large (several effective radii) \citep[see also,][]{Churazov.et.al.2010}. Also a single power-law model  seems to be a poor approximation of the total mass density distribution at radii $\lesssim R_{\mathrm{eff}}$ especially for less massive galaxies with $\sigma_{ap}(<R_{\mathrm{eff}}/2) \lesssim 200-250$ km s$^{-1}$.

Despite possible systematic biases and/or difficulties with the interpretation of the estimated $\gamma$, the Koopmans et al. procedure has proved to be a powerful tool for analysing large samples of lenses (SLACS, SL2S) with available aperture velocity dispersion measurements and also a way of comparing simulated galaxies with real ones \citep[see, e.g.][]{Xu.et.al.2017}.

Compared to the Koopmans et al. approach, the method used here is non-parametric and requires the  velocity dispersion profile rather than a single aperture velocity dispersion measurement. This additional information allows one to avoid the difficulty  with the interpretation  of the estimated $\gamma$ and partially reduce the uncertainty associated with the anisotropy. All in all the scatter in individual measurements of the mass slopes of massive elliptical galaxies is reduced to $\simeq 13\%$ according to our tests on simulated galaxies.

\section{Real galaxies}
\label{sec:lensing_sample}

\subsection{Sample description}

In the Section~\ref{seq:slope} we have tested our  approach for density slope estimation on simulated elliptical galaxies. Let us now turn to real objects, namely, to a sample of 15 SLACS\footnote{the Sloan Lens ACS Survey} lens galaxies at $0.01 \leq z \leq 0.35$ (Table~\ref{tab:table}). 
These early-type galaxies have already been analysed in detail in \cite{Barnabe.et.al.2009, Barnabe.et.al.2011} using the state-of-the-art joint lensing and kinematic code CAULDRON\footnote{`Combined Algorithm for Unified Lensing and Dynamics ReconstructiON' \citep{Barnabe.Koopmans.2007,Barnabe.et.al.2009,Barnabe.et.al.2011}}.
The obtained simple circular speed estimates ($V_c(R_{opt})=\sqrt{GM(r=R_{opt})/R_{opt}}$) using equations~(\ref{eq:Ropt}) and (\ref{eq:VcR2}) in comparison with the circular speed curves from the detailed analysis of \cite{Barnabe.et.al.2009, Barnabe.et.al.2011} are presented in Section~\ref{appendix}, Figure~\ref{fig:galaxies}.
On average, our $V_c$-estimates agree well with the results of the more sophisticated analysis from \cite{Barnabe.et.al.2009, Barnabe.et.al.2011}  despite all the simplifying assumptions.

\begin{table*}
\centering
\caption{\label{tab:table} Basic properties of the 15 SLACS lens galaxies. The data on the lens redshifts $z_{lens}$, the SDSS velocity dispersions $\sigma_{\mathrm{SDSS}}$ (uncorrected for aperture effects), and effective radii $R_{\mathrm{eff}}$ in V-band are taken from \citealt{Bolton.et.al.2008}. Einstein radii $R_{Einst}$, Einstein masses $M_{Einst}$, and effective radii in I-band come from \citealt{Auger.et.al.2009}.}
\begin{tabular}{lccccccccc}
\hline
(1)  & (2)     & (3)          & (4)                        & (5)         & (6)  & (7) \\       
name & $z_{lens}$ & $R_{Einst}$, kpc & $M_{Einst}, 10^{11} M_{\odot}$ & $R_{\mathrm{eff}}$, kpc (V-band)&  $R_{\mathrm{eff}}$, kpc (I-band)& $\sigma_{\mathrm{SDSS}}$, km s$^{-1}$\\
\hline
SDSS J0037--−0942 & 0.1955 & 4.95 & 2.95 & 7.11  & 5.84  & $279 \pm 14$\\
SDSS J0216--0813  & 0.3317 & 5.53 & 4.90 & 12.74 & 11.46 & $333 \pm 23$\\
SDSS J0912+0029   & 0.1642 & 4.58 & 3.98 & 10.90 & 11.30 & $326 \pm 16$\\
SDSS J0935--0003  & 0.3475 & 4.26 & 3.98 & 20.85 & 10.57 & $396 \pm 35$\\
SDSS J0959+0410   & 0.1260 & 2.24 & 0.76 & 3.14  & 2.91  & $197 \pm 13$\\
SDSS J1204+0358   & 0.1644 & 3.68 & 1.74 & 4.15  & 3.07  & $267 \pm 17$\\
SDSS J1250+0523   & 0.2318 & 4.18 & 1.82 & 6.69  & 4.88  & $252 \pm 14$\\
SDSS J1330--0148  & 0.0808 & 1.32 & 0.33 & 1.36  & 1.46  & $185 \pm 9$\\
SDSS J1443+0304   & 0.1338 & 1.93 & 0.60 & 2.23  & 1.66  & $209 \pm 11$\\ 
SDSS J1451--0239  & 0.1254 & 2.33 & 0.88 & 5.58  & 3.46  & $223 \pm 14$\\
SDSS J1627--0053  & 0.2076 & 4.18 & 2.29 & 6.73  & 6.73  & $290 \pm 15$\\
SDSS J2238--0754  & 0.1371 & 3.08 & 1.29 & 5.65  & 4.41  & $198 \pm 11$\\
SDSS J2300+0022   & 0.2285 & 4.51 & 2.95 & 6.70  & 5.56  & $279 \pm 17$\\
SDSS J2303+1422   & 0.1553 & 4.35 & 2.63 & 8.83  & 7.92  & $255 \pm 16$\\
SDSS J2321--0939  & 0.0819 & 2.47 & 1.20 & 6.35  & 6.35  & $249 \pm 12$\\
\hline
\end{tabular}
\label{tab:sample}
\end{table*}

\subsection{Slope of the density profile}
\label{sec:slope}

Given the Einstein masses and  mass estimates from kinematic analysis, we can evaluate the density slope of the lens galaxies under  the assumptions that galaxies are spherically symmetric systems with power-law density distribution.
 We determine the density slope as the peak of the $P(\gamma)$-distribution calculated as

\be
\displaystyle P(\gamma)\propto \exp\left({-\frac{1}{2}\frac{(V_c(r,\gamma)-V_0(R_{opt}))^2}{(\Delta V_0)^2}}\right),
\ee
where $V_c(r,\gamma)$ comes from strong lensing measurements (equation~(\ref{eq:Vcgamma})), and $V_0(R_{opt})\pm \Delta V_0 $ is the dynamical $V_c$-estimate at some optimal radius $R_{opt}$. Uncertainties on $\gamma$ (arising from kinematic uncertainties only) are calculated as a 68\% confidence interval. 

In the Section~\ref{sec:method} we have considered two  dynamical mass estimators:  $M(0.5 R_{\mathrm{eff}})$ and $M(R_2)$. Both of them are demonstrated to agree well with results of more sophisticated analyses. For the purpose of estimating the density slope we focus on the $V_0(R_{opt})=V_c(0.5R_{\mathrm{eff}})$ estimator for several reasons.
First of all, the surface brightness distribution of the galaxies under consideration is generally well fitted with the de Vaucouleurs profile so the  $V_c$-estimate at $\simeq 0.5R_{\mathrm{eff}}$ is expected to be minimally sensitive to the anisotropy \citep{Churazov.et.al.2010}. Secondly, 
observational uncertainties at $0.5R_{\mathrm{eff}}$ are typically considerably smaller than at $R_2 \simeq R_{\mathrm{eff}}$. And finally, for nearby lens galaxies $0.5R_{\mathrm{eff}}$ is generally smaller than $2.2R_{Einst}$, and $P(\gamma)$ has a well-defined peak. While in the case $R_{opt}=R_2 \gtrsim 2.2R_{Einst}$ the probability density of $\gamma$ is likely to exhibit a broad peak or to be two-peaked. 

\begin{figure*}
\plotone{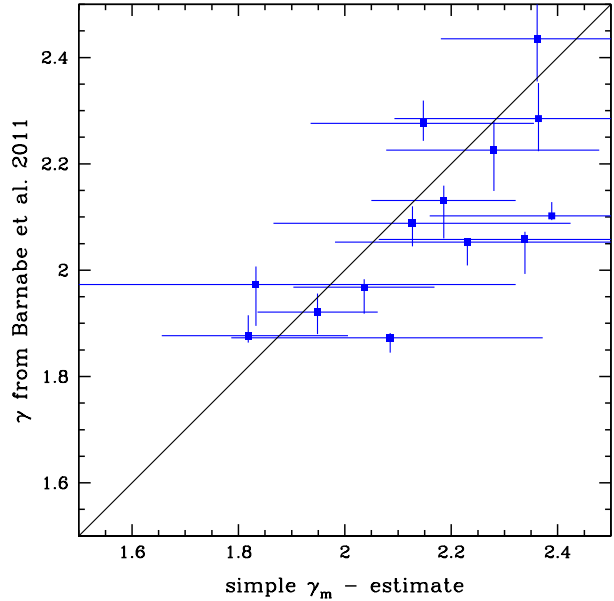}  
\caption{Our estimates of the (mass-weighted) logarithmic slope $\gamma$ of the total density profiles $\rho \propto r^{-\gamma}$ in comparison with detailed modelling from \citealt{Barnabe.et.al.2011} which self-consistently combines lensing and dynamics constraints. Error bars (indicating a 68\% confidence interval) of our estimates are conservative and come from observational uncertainties in kinematic profiles.
  Density exponents $\gamma$ from  \citealt{Barnabe.et.al.2011} are shown with 99\% confidence error bars  calculated by considering the region of the marginalized posterior probability density function for $\gamma$.
\label{fig:slope}
}
\end{figure*}

Figure~\ref{fig:slope} shows the resulting estimates of the density exponent (between $0.5R_{\mathrm{eff}}$ and $2.2R_{Einst}$) in comparison with values reported in \cite{Barnabe.et.al.2011}. However, one should bear in mind that a comparison between our slope estimates and ones from \cite{Barnabe.et.al.2011} is not straightforward. The former is the average slope of the mass profile between two `pinch' radii, while the latter is the best-fit of an axisymmetric power-law profile to the total density. Moreover, the CAULDRON requires that the surface brightness distribution of the lensed background galaxy must be reproduced by the lens model, what basically means that more weight is  assigned to lensing data rather than kinematics  (as opposed to our simple approach). In \cite{2009MNRAS.393.1114B} it has been shown that CAULDRON recovers reasonably well the mass slope between $\simeq R_{\mathrm{eff}}/3-R_{\mathrm{eff}}/2 $ and $\simeq R_{\mathrm{eff}}$ for not too face-on galaxies. So both approaches recover mass slopes over roughly the same radial range.  So we do not expect neither the perfect agreement between these two types of $\gamma$, nor the significant disagreement. The rms-scatter between different estimates is $\simeq 15\%$. 

The sample-averaged logarithmic density slope found from the simple circular speed estimates at $R_{opt} \simeq 0.5R_{\mathrm{eff}}$ and the measurement of (projected) Einstein masses $M_{Einst}$ is $\left< \gamma \right> = 2.1\pm 0.05$ (68\%CL).

\section{Conclusions and discussion}
\label{seq:disc}

Although neither dark matter nor stars in early-type galaxies have a power-law density distribution, the total mass density is reasonably well described by a single power-law $\rho \propto r^{-\gamma}$. This has been shown in numerous observational and numerical modelling studies   \citep[e.g.,][among others]{Remus.et.al.2013, Cappellari.et.al.2015}. 
Total mass density slopes and their evolution with redshift might contain significant information about the physical processes during the evolution of massive galaxies as they are sensitive to stellar and AGN feedback processes \citep[see, e.g.][]{Remus.et.al.2017}.
In this work we have discussed easy-to-use and fast in implementation approaches for recovering masses and mass slopes of massive early-type galaxies by combining the Einstein masses/radii measurements from strong lensing observations with information on the stellar kinematics of lens galaxies.
If a spatially resolved velocity dispersion is available up to $\sim (0.5 - 1) R_{\mathrm{eff}}$  then one can find a special (`optimal') radius, at which the mass-estimate of a galaxy is minimally affected by details of the unknown anisotropy profile. On the other hand, given an estimate of the Einstein radius and the aperture mass interior to it, an independent (from dynamics) mass probe at another (in general) special radius, where sensitivity to the density model is expected to be minimal, can be recovered. Two masses at different radii trivially give the mass slope (or mass-weighted density slope) between these radii. We have tested this procedure on a set of simulated galaxies from the high-resolution cosmological zoom simulations of \cite{Oser.et.al.2010, Oser.et.al.2012} and compared it to the widely used approach of      
\cite{Koopmans.et.al.2006}. The main results are the following:

\begin{itemize}
\item Under an assumption of spherical symmetry and a power-law total density distribution $\rho \propto r^{-\gamma}$, the mass of a galaxy at $\simeq 2.2 R_{Einst}$  is given by
  $$M(2.2 R_{Einst}) \simeq \eta(\gamma) M_{Einst},$$  where $\eta$ varies from 1.23 to 1.4 when $\gamma$ changes from $1.5$ to $2.5$. 
 This mass estimate is based only on the measurements of Einstein radius $R_{Einst}$ and the Einstein mass $ M_{Einst}$, and  is largely independent of the details of the total density distribution. Tests on the simulated galaxies have shown that this simple formula  works well for majority of relatively massive galaxies with $ \sigma_{ap}(<R_{\mathrm{eff}}/2)  \gtrsim 200 $ km $s^{-1}$, where $\sigma_{ap}(<R_{\mathrm{eff}}/2)$ is the aperture velocity dispersion measured within one-half of the effective radius.

\item  The mass slopes of simulated galaxies between $2.2 R_{Einst}$ and the optimal kinematic radius $R_{opt}$, at which the mass estimate is  largely  insensitive to the anisotropy, are recovered with     $\simeq 13$ per cent accuracy (rms-scatter).
The main uncertainty-driver is the ratio between $R_{opt}$ and $2.2 R_{Einst}$. If these two radii are two close to each other then the estimated mass slope is not unique or has large error bars.
 While the accuracy of this approach has been tested on a sample of (relatively isolated) simulated massive galaxies, a sample of real elliptical galaxies acting as strong lenses could imply an additional selection bias that can potentially introduce a non-trivial bias in the estimated mass slope.

\item  We have applied our simple technique to derive the mass-weighted density slopes of 15 SLACS lens galaxies   at $0.01 \le z \le 0.35$ with spatially resolved kinematics.   These early-type galaxies have already been analysed in detail in \cite{Barnabe.et.al.2009, Barnabe.et.al.2011} by state-of-the-art joint lensing and kinematic code CAULDRON  \citep{Barnabe.Koopmans.2007,Barnabe.et.al.2009,Barnabe.et.al.2011}. Both the mass normalizations and the mass slopes are in general in a good agreement with results of more sophisticated modelling. Although for 2 galaxies in the studied sample (namely, J1627 and J2303) the estimated slopes are noticeably steeper than obtained in \cite{Barnabe.et.al.2011}  but within the errors. These galaxies are massive, roundish and slowly rotating so we do not expect  significant systematic biases in dynamical circular speed (or mass) estimation at the optimal radius $R_{opt}$. Moreover, recent high signal-to-noise long-slit velocity dispersion and rotation velocity measurements of J1627, obtained within the framework of  the X-shooter Lens Survey (XLENS) \cite{Spiniello.et.al.2015}, strengthen this tension as the XLENS velocity dispersion for J1627 is  higher than VIMOS one. This leads to even higher circular speed estimate at  $R_{opt}$ and, in principle, steeper slope.   

\end{itemize}

Although the discussed simple method is intrinsically much less comprehensive than advanced lensing/dynamical modelling, it is fast in implementation and uses only few arithmetic operations. 
The general agreement  with results from simulations and the state-of-the-art analysis performed by  \cite{Barnabe.et.al.2009, Barnabe.et.al.2011} makes it a reliable means to estimate masses and mass slopes of massive elliptical galaxies by manipulating  the surface brightness profile, spatially resolved velocity dispersion and Einstein mass/radius measurements.

\section{Acknowledgments} 
This work was partially supported by the Russian Science Foundation (grant 14-22-00271).
TN acknowledges support from the DFG cluster of excellence `Origin and Structure of the Universe'.
NL is thankful to Tomasso Treu and Matteo Barnab{\`e} for useful discussions and observational data, and Chiara Spiniello for providing the XLENS data. We are grateful to the referee for useful comments and suggestions which helped to improve the paper.

\vspace{1cm}

\appendix

\section{}

\label{appendix}

\begin{figure*}
\plottwos{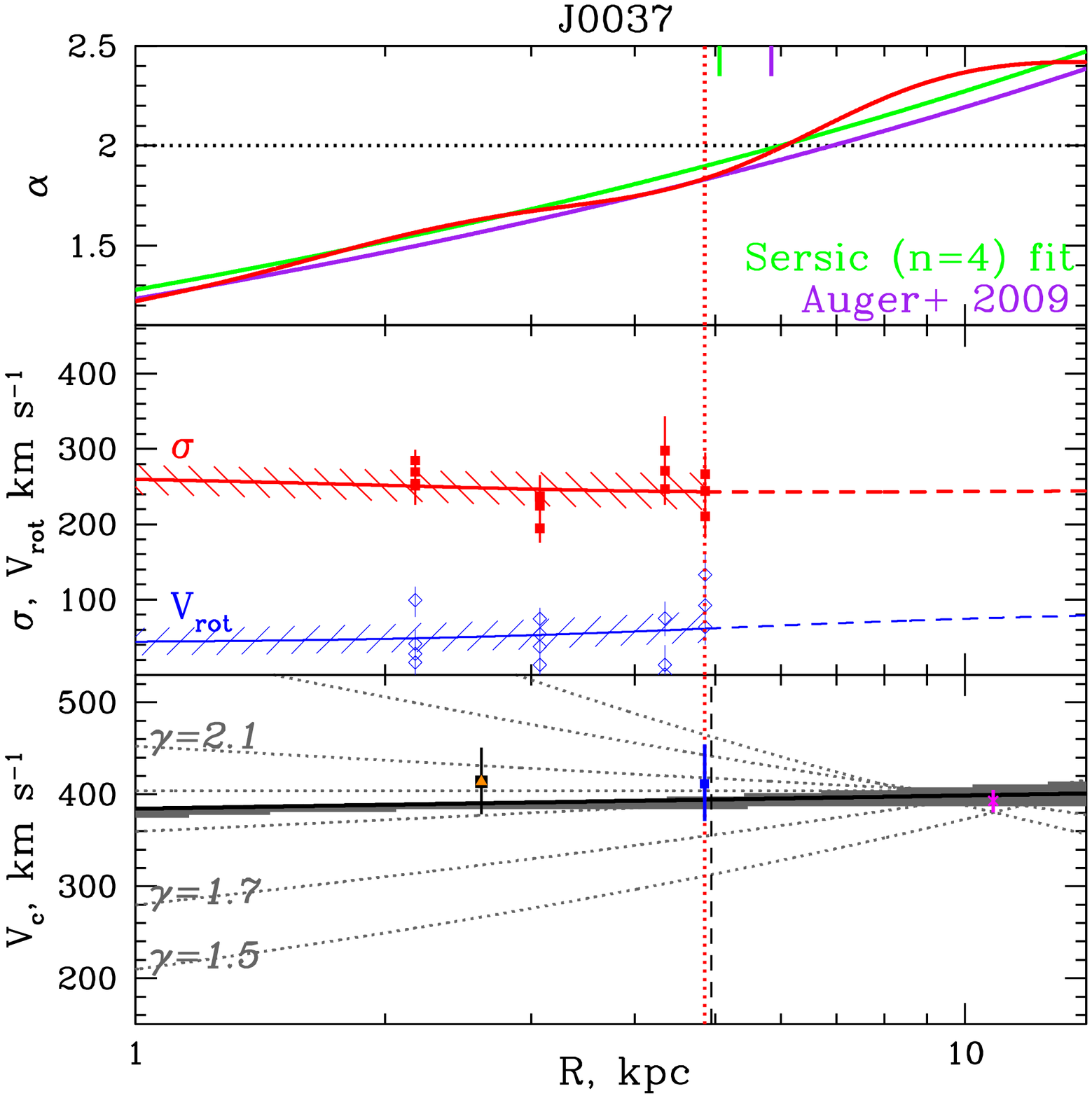}{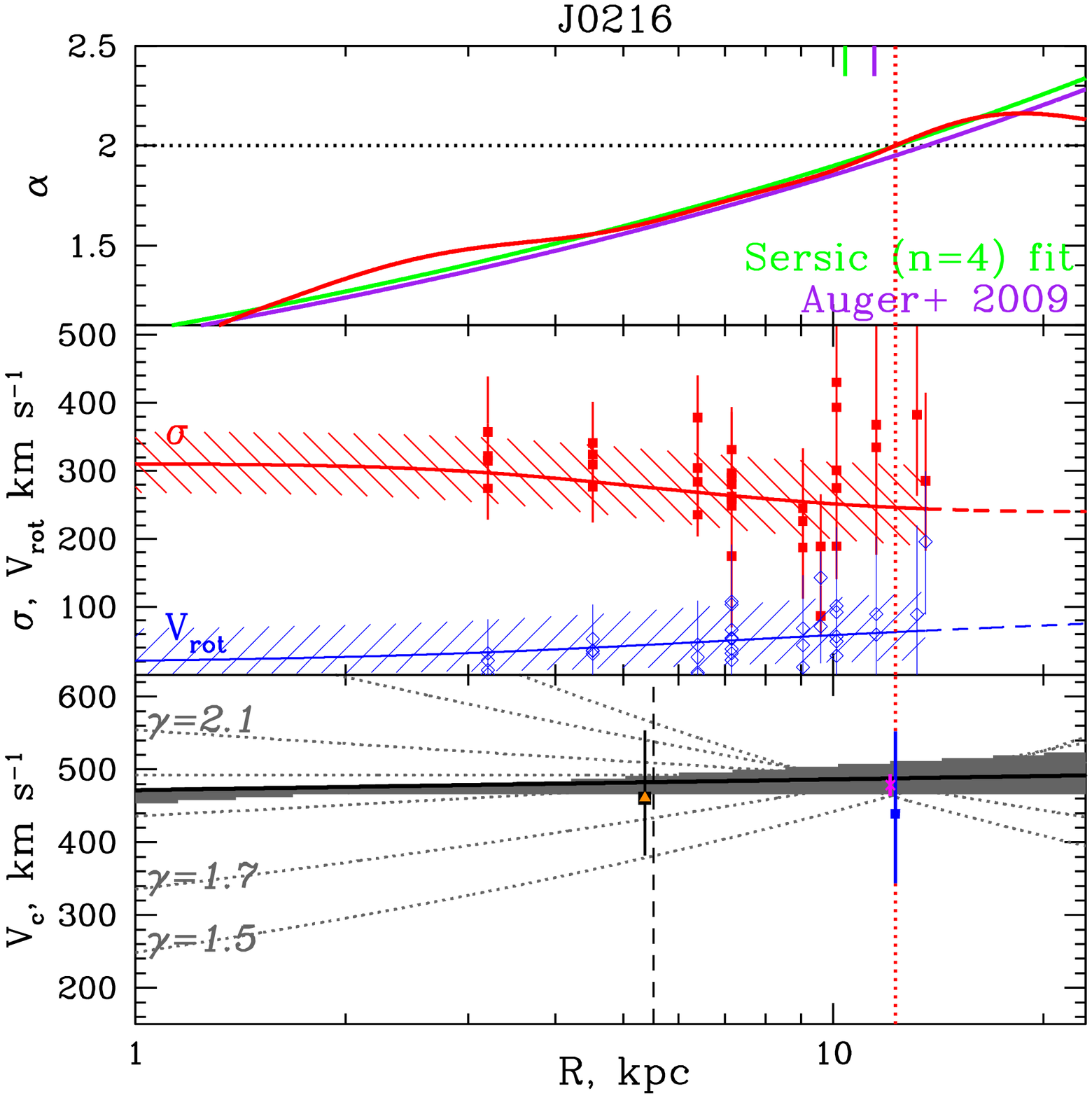}
\plottwos{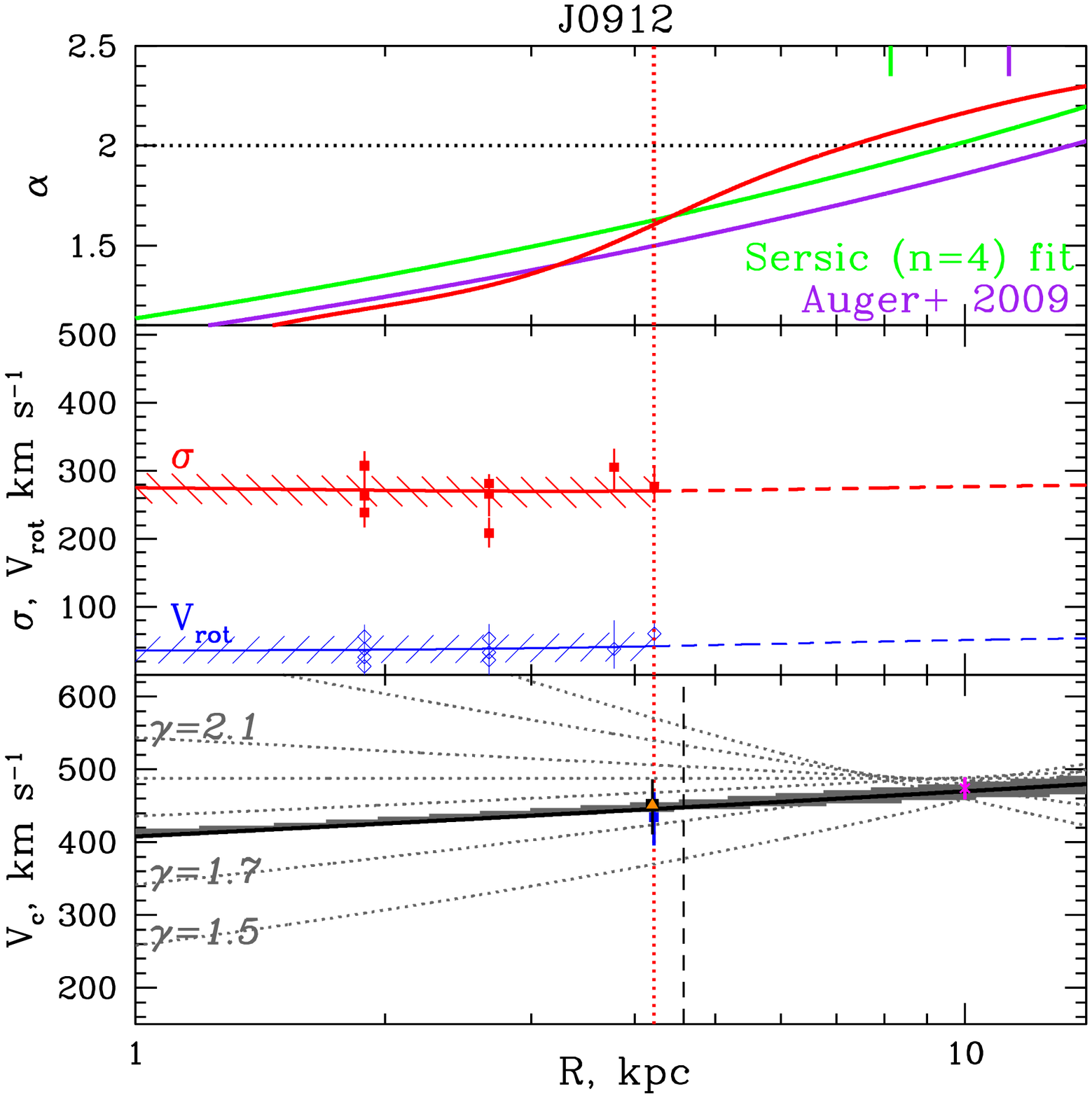}{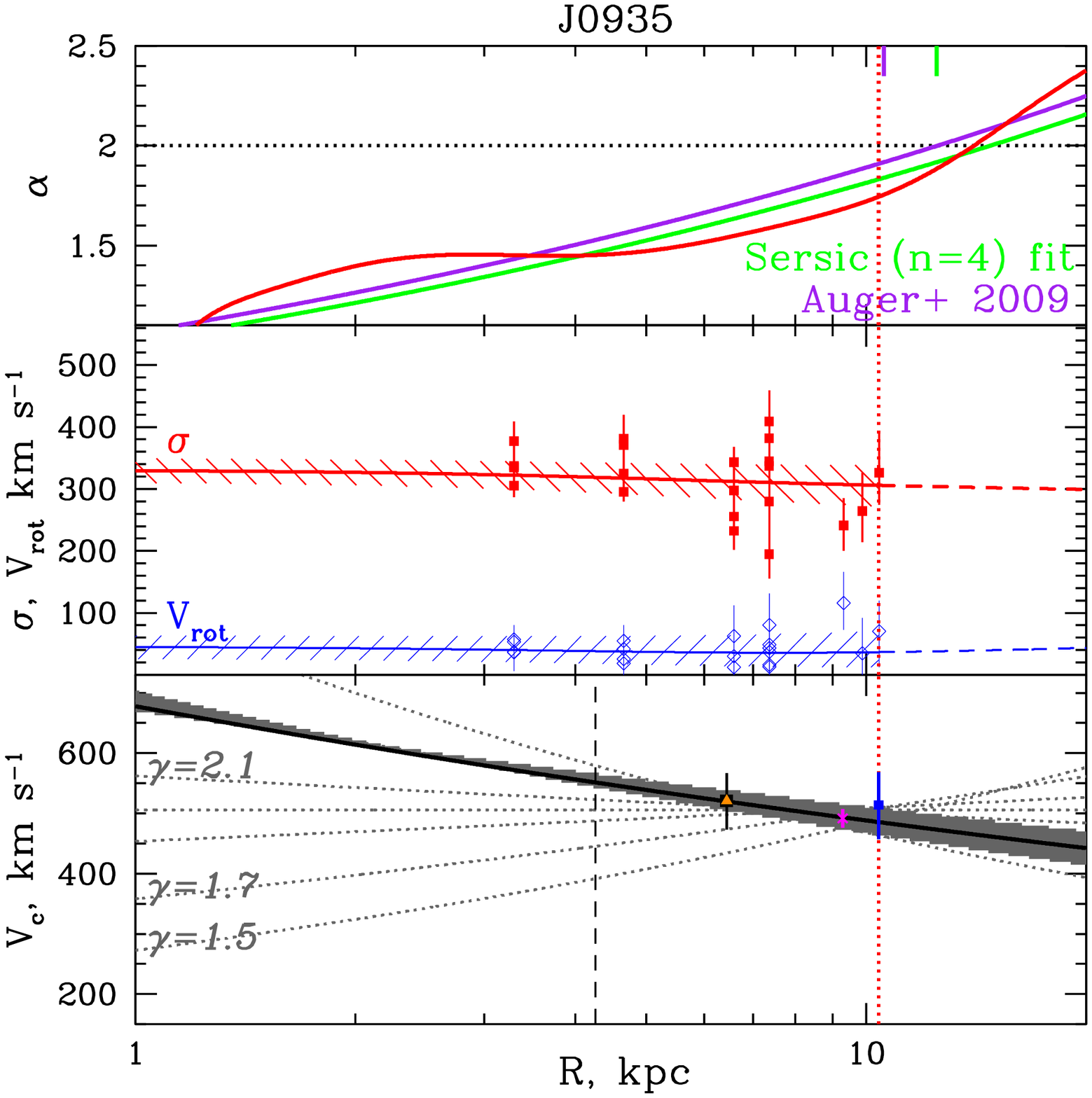}
\plottwos{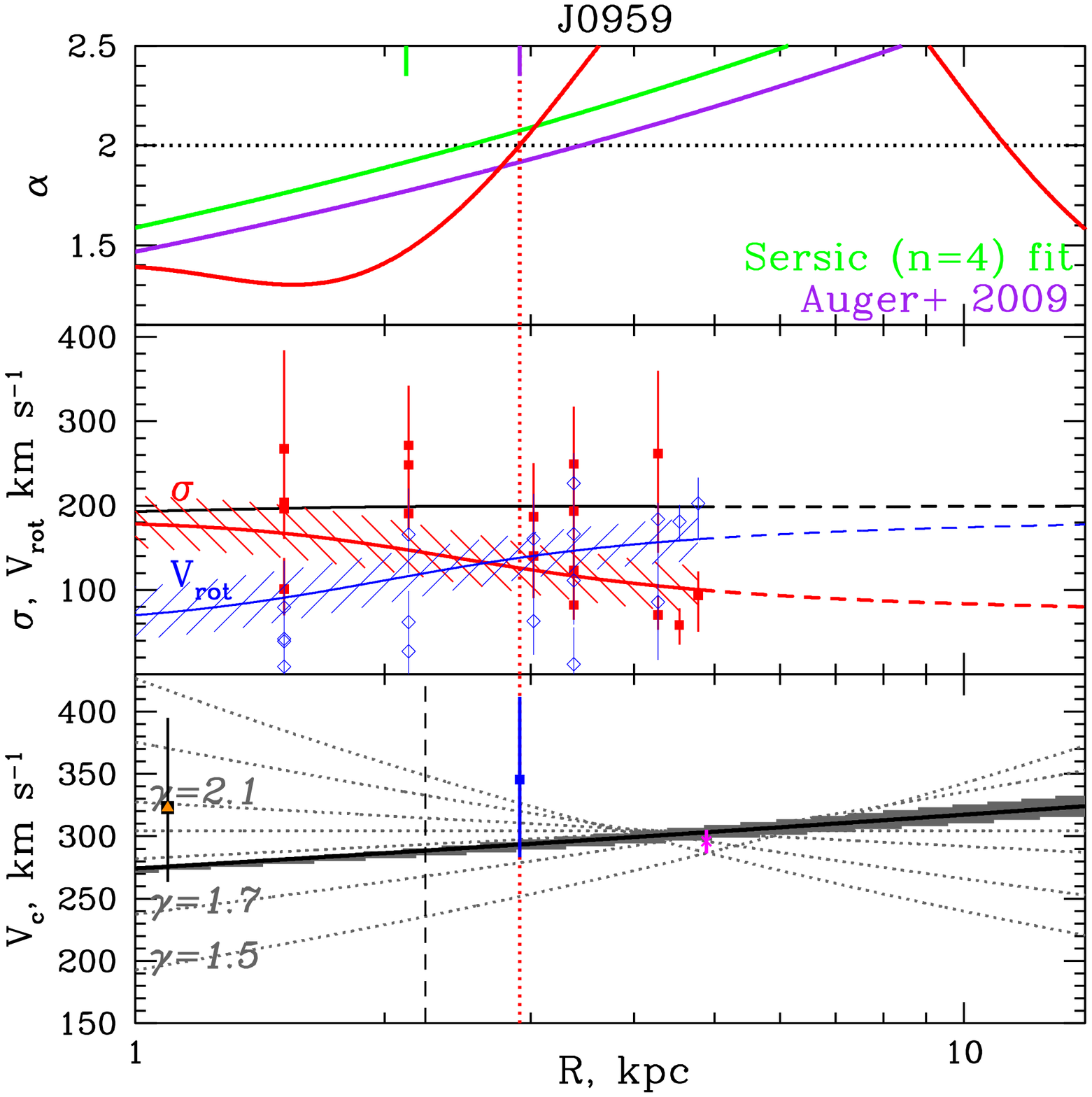}{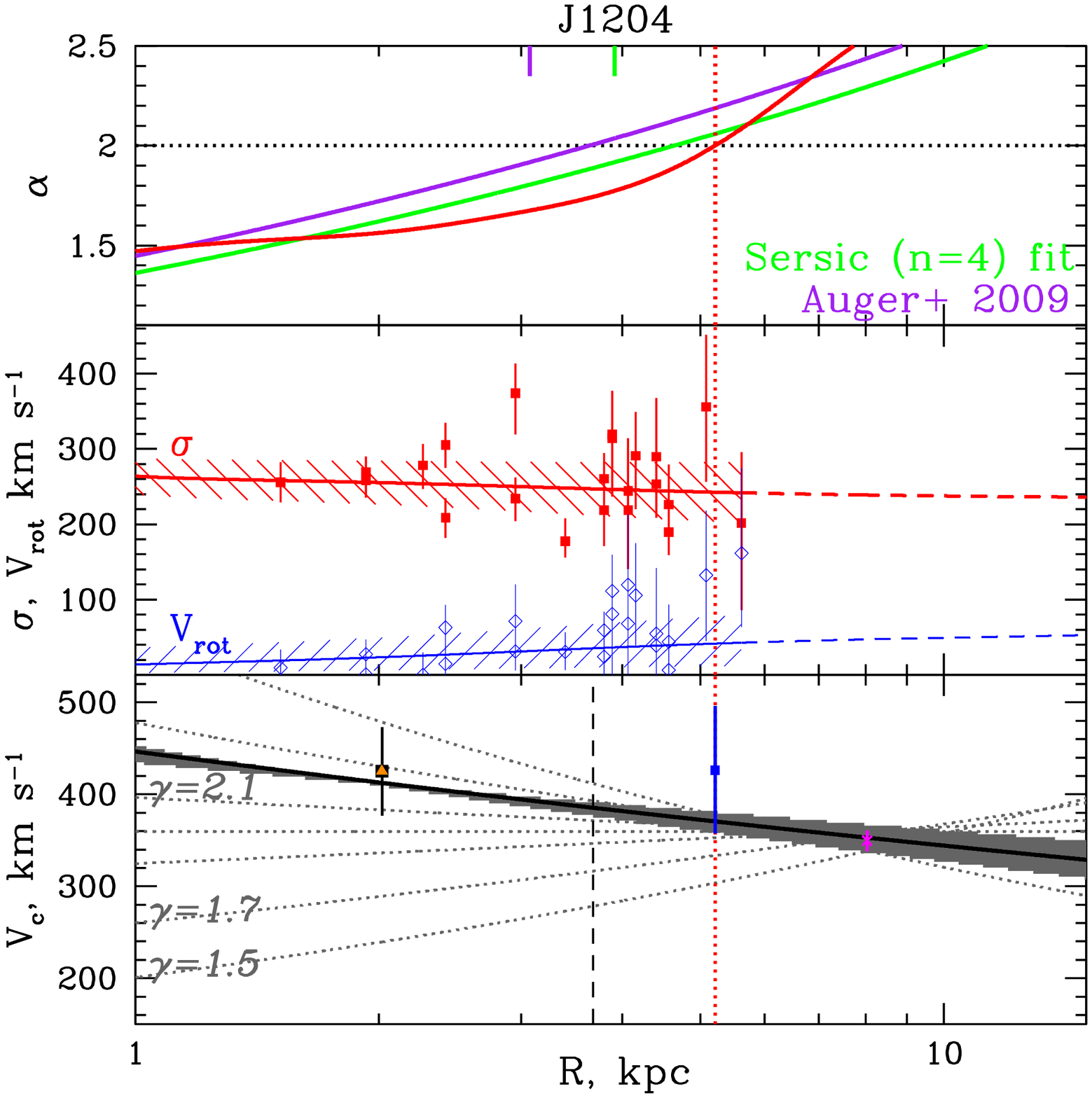}
\caption{Comparison of simple $V_c$-estimates with the circular speed from \citealt{Barnabe.et.al.2009, Barnabe.et.al.2011} obtained via lensing and dynamics code CAULDRON.
Upper panels show the logarithmic slopes $\alpha = -d \log I/d \log R$ of the surface brightness profile (in red) extracted in circular annuli and the slope of its de Vaucouleurs fit (in green), and the slope of the 2D ellipsoidal de Vaucouleurs fit to the image from \citealt{Auger.et.al.2009}.  The corresponding effective radii are marked with green and purple line segments.
Middle panels show the projected velocity dispersion $ \sigma$ and rotation velocity $ V_{rot}$ ( and RMS velocity ($ V_{RMS}=\sqrt{\sigma^2+V_{rot}^2}$)  for rapidly rotating galaxies)  as well as interpolated curves. Kinematic profiles are calculated in circular annuli. 
Lower panels  present the circular speed resulting from the joint lensing and kinematics analysis (black thick curve) with error bars (grey shaded region) and the simple $V_c$-estimates. The black dashed line and the red dotted line show the Einstein radius and $R_2$ respectively.
\label{fig:galaxies}
}
\end{figure*}

In this Appendix we present our
simple circular speed estimates ($V_c(R_{opt})=\sqrt{GM(R_{opt})/R_{opt}}$ at $0.5R_{\mathrm{eff}}$ and at $R_2$) in comparison with the circular speed curves from the detailed analysis of \cite{Barnabe.et.al.2009, Barnabe.et.al.2011} (Figure~\ref{fig:galaxies}). The upper panels in Figure~\ref{fig:galaxies} show the logarithmic slope $\displaystyle \alpha=-d\log I/d\log R$ of the surface brightness profile (red curve), extracted in circular annuli, and its de Vaucouleurs fit (in green). The corresponding effective radius is marked with a short green line segment on the top of the upper panels. For comparison we also plot the slope and mark the effective radius\footnote{defined in \cite{Auger.et.al.2009} as the geometrical mean of the major and minor axes of the elliptical isophotal contour enclosing one-half of the model flux} of the de Vaucouleurs 2D ellipsoidal fit (in purple) provided in \cite{Auger.et.al.2009}.    
The middle panels show data (in circular annuli) on the line-of-sight velocity dispersion (in red) and the line-of-sight velocity (in blue) obtained with the integral-field unit of VIMOS on the VLT as well as smoothed (interpolated) curves with  uncertainties.
The full description of the used observational data  is given  in \cite{Czoske.et.al.2012}.
For the rapidly rotating systems (J0959, J1330, and J2238) the interpolated curve for the RMS velocity $ V_{RMS}=\sqrt{\sigma^2+V_{rot}^2}$ is given  and $V_{RMS}$ at $R_{opt}$ is used instead of $\sigma(R_{opt})$ as a proxy for the circular velocity, i.e. $V_{RMS}(R_{opt}) \simeq 0.6V_c(R_{opt})$ \citep{Lyskova.et.al.2014}. In the calculation of $V_{RMS}$ we assume that no information on an inclination angle of the rotation axis with respect to the line-of-sight is given. 
The lower panels present the circular speed curves from \cite{Barnabe.et.al.2009, Barnabe.et.al.2011} (black thick lines) with uncertainties and the simple circular speed estimates at $R_{opt} \simeq 0.5R_{\mathrm{eff}}$ (orange symbol) and at $R_2$ (blue symbol). The errorbars for the simple $V_c$-estimates mainly come from the observational uncertainties in $\sigma$-profile (or in $V_{RMS}$-profile for rapidly rotating galaxies). For the systems with $R_2$ lying beyond the outermost boundary $R_{kin}$ of the kinematic data we estimate the circular speed at $R_{kin}$. Estimated errors arising from difference between $V_c^{iso}$, $V_c^{circ}$, and $V_c^{rad}$ are typically smaller than observational kinematical uncertainties. $R_{Einst}$ and $R_2$ (or the outermost radius probed by the kinematic data set) are marked with a dashed and dotted lines correspondingly. Also we show predicted $V_c$-curves for a $\rho \propto r^{-\gamma}$ model (equation~\ref{eq:Mgamma}) given the $M_{Einst}$ -values from Table~\ref{tab:sample}. On average, simple $V_c$-estimates agree well with the circular speed curves derived in \cite{Barnabe.et.al.2009, Barnabe.et.al.2011} despite all the simplifications.

\setcounter{figure}{0}
\begin{figure*}
\plottwos{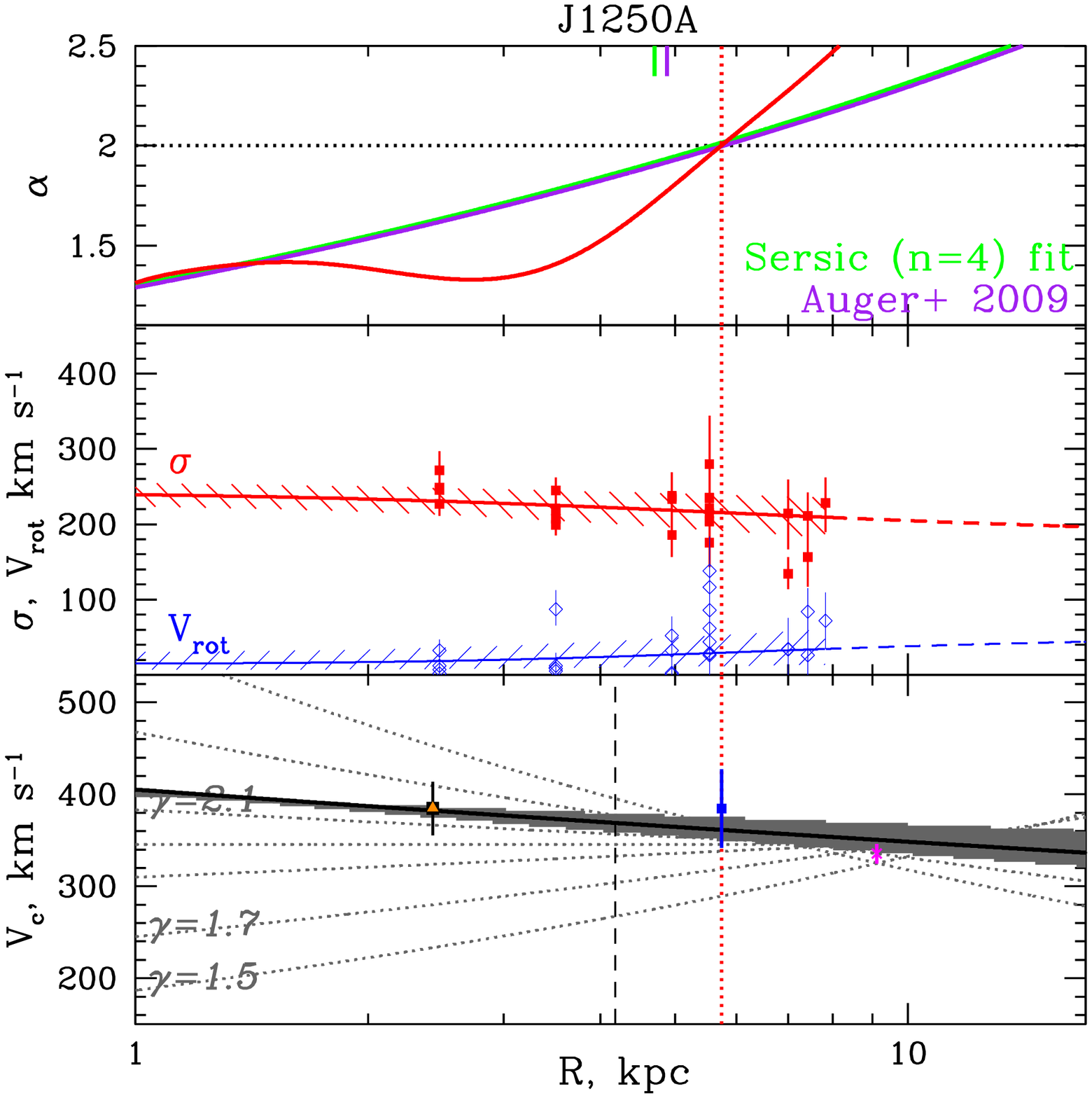}{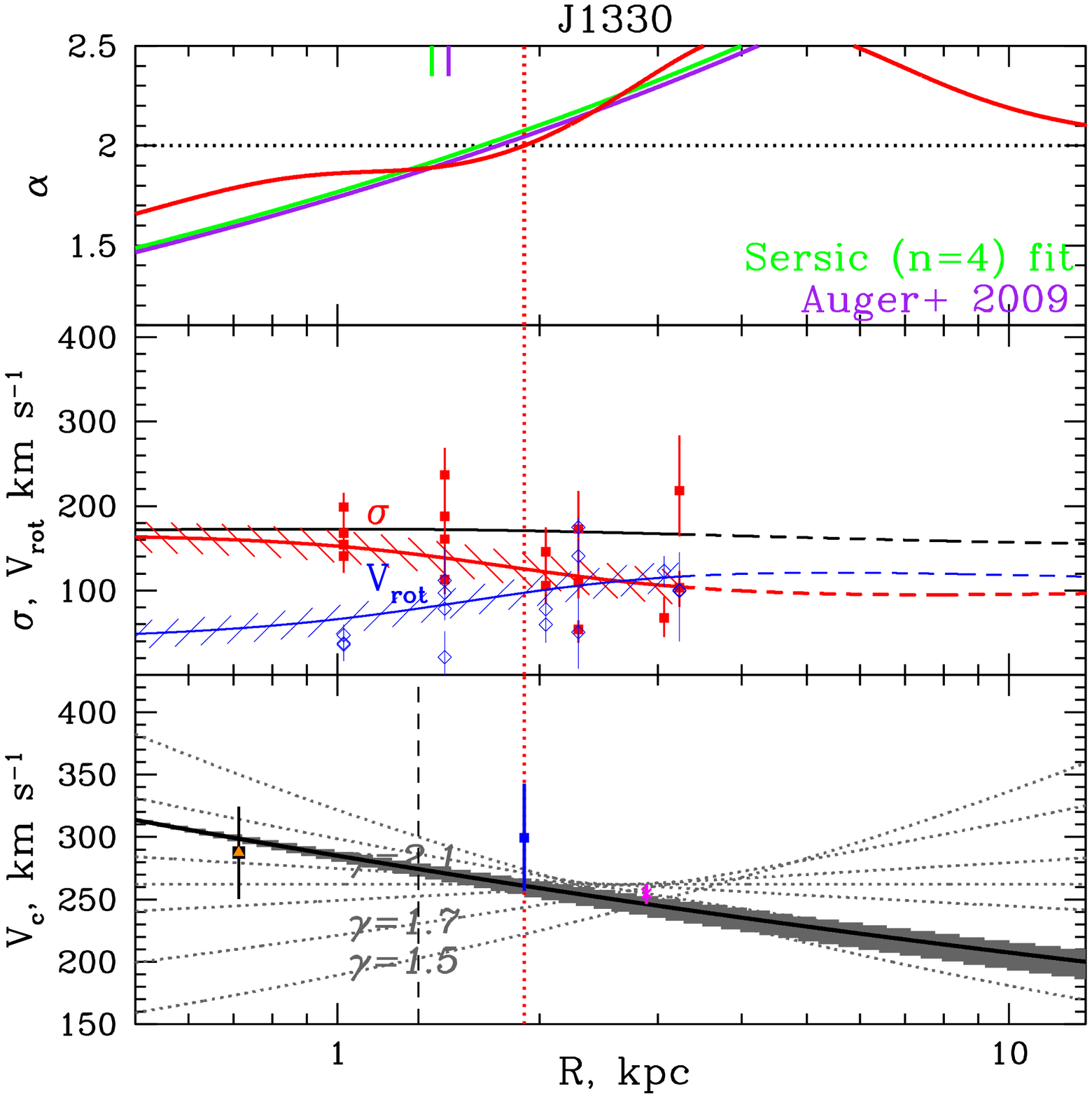}
\plottwos{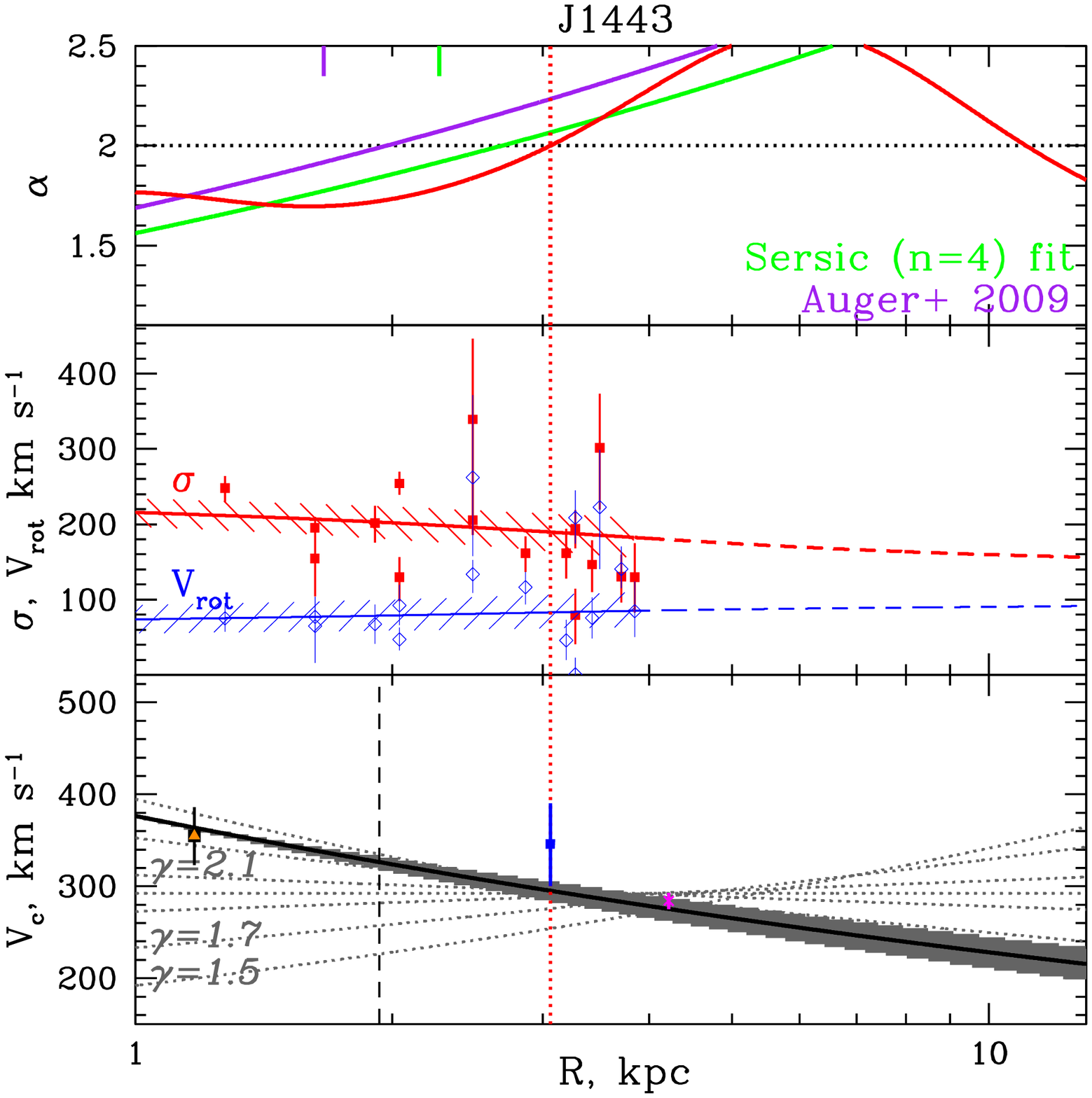}{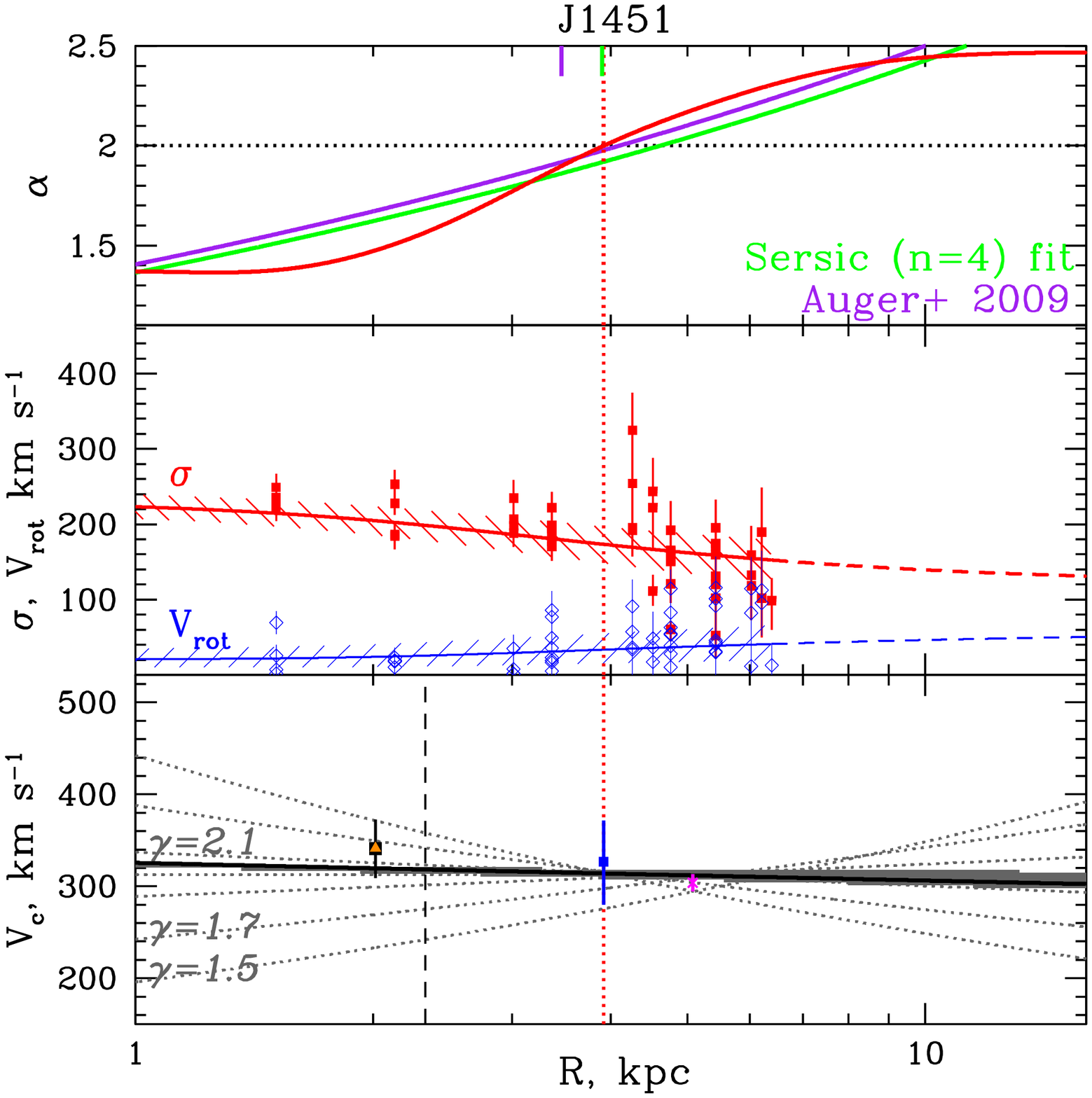}
\plottwos{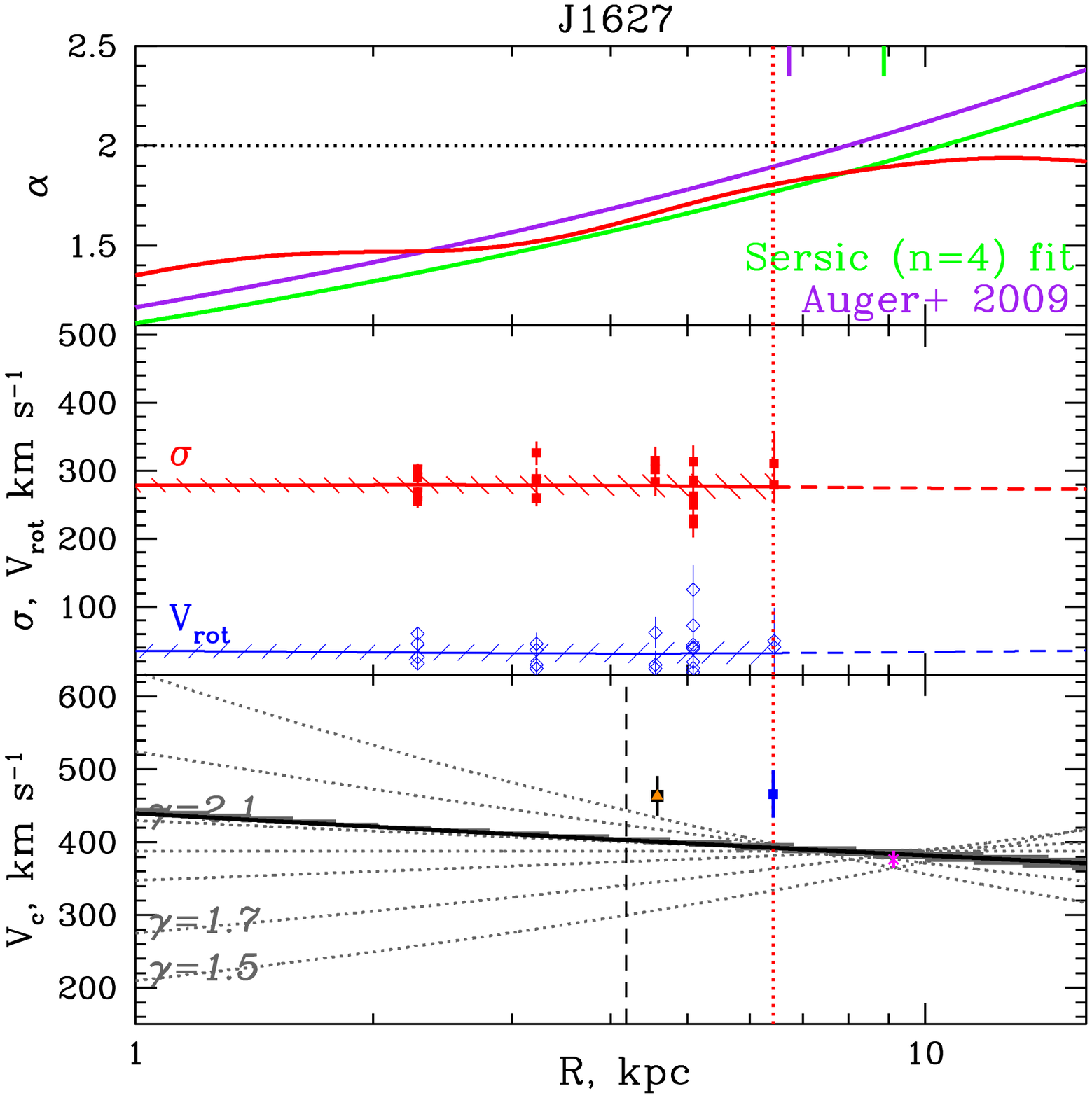}{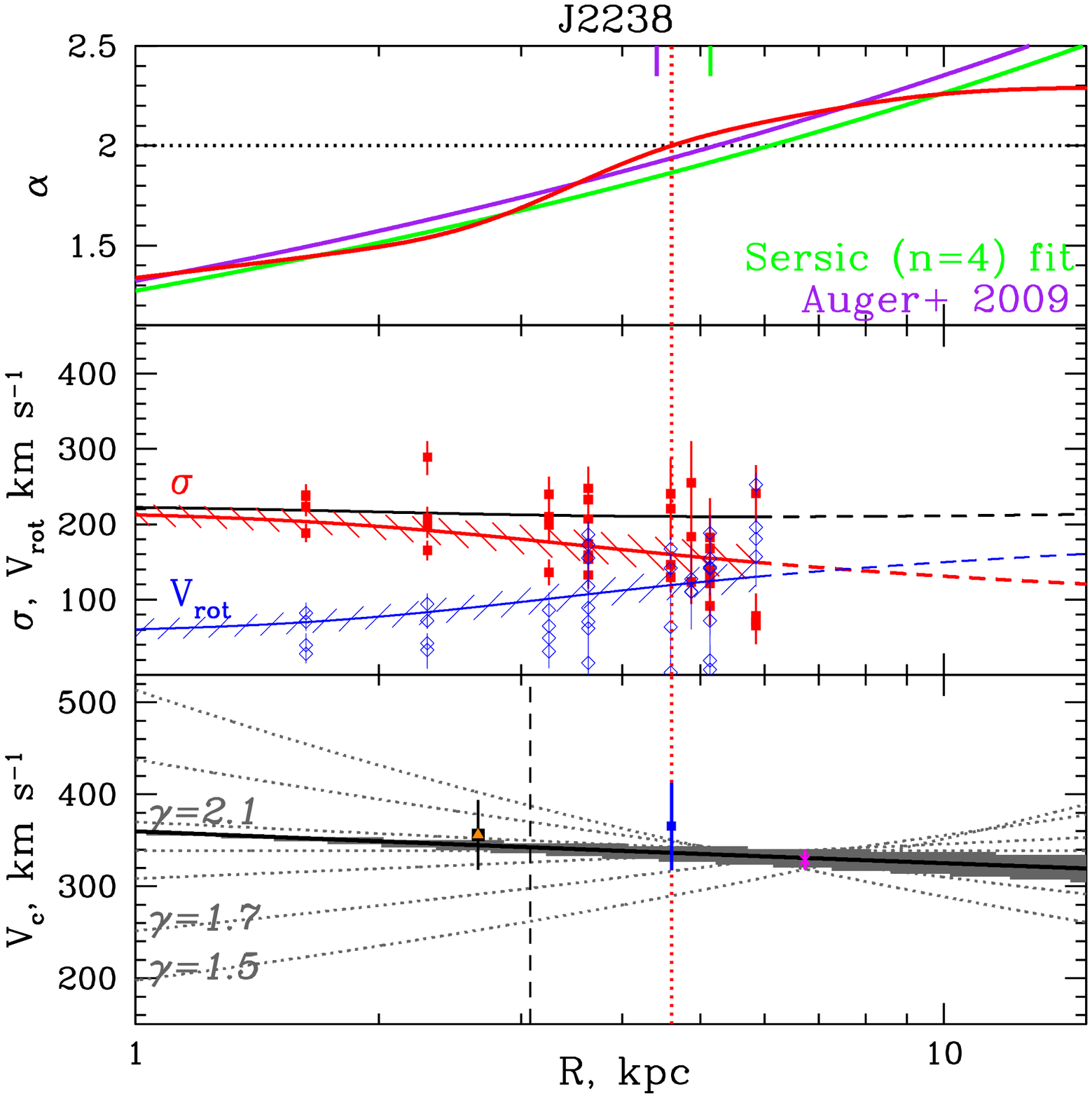}
\caption{(continue)} 
\end{figure*}

\setcounter{figure}{0}
\begin{figure*}
\plottwos{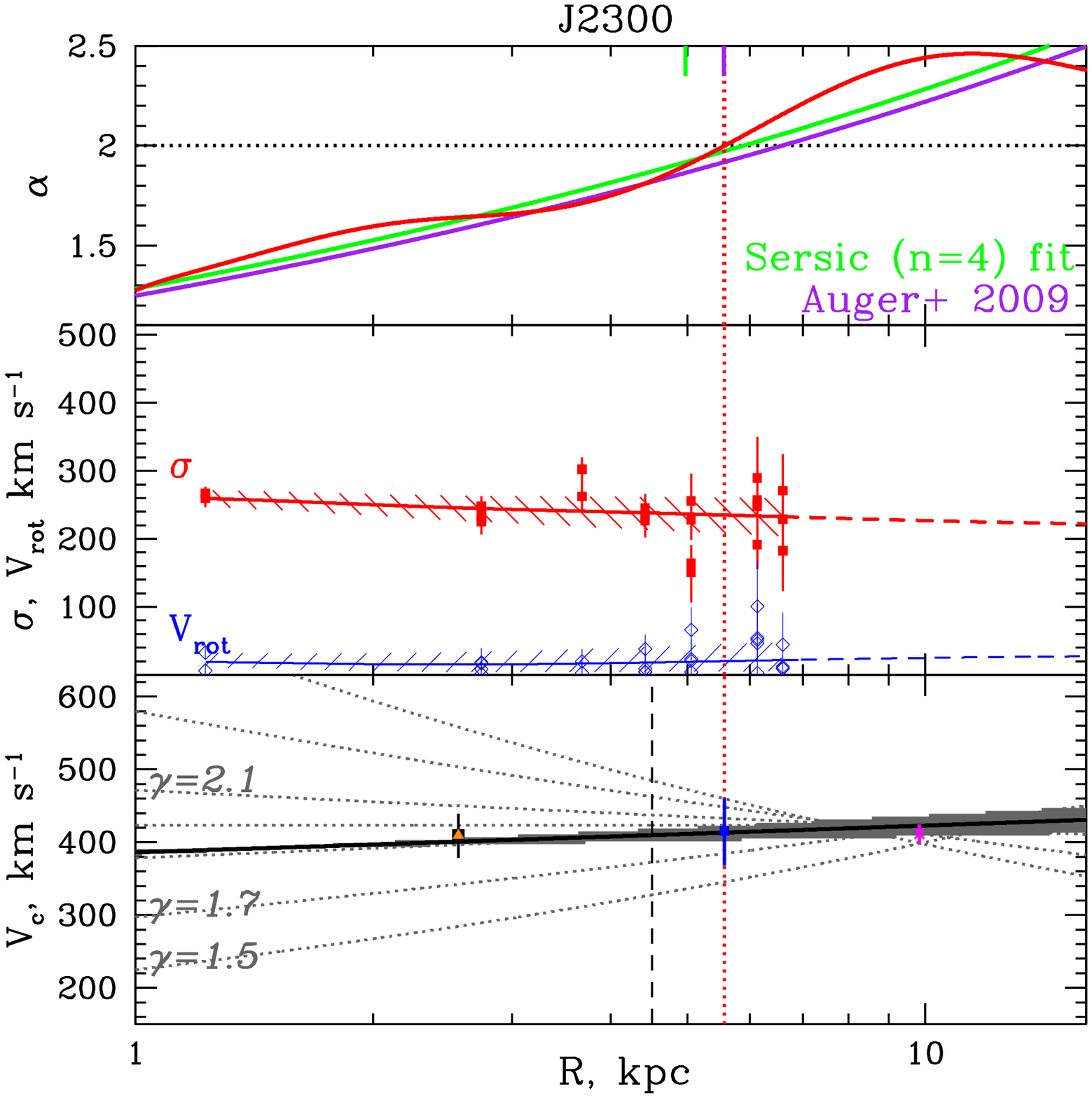}{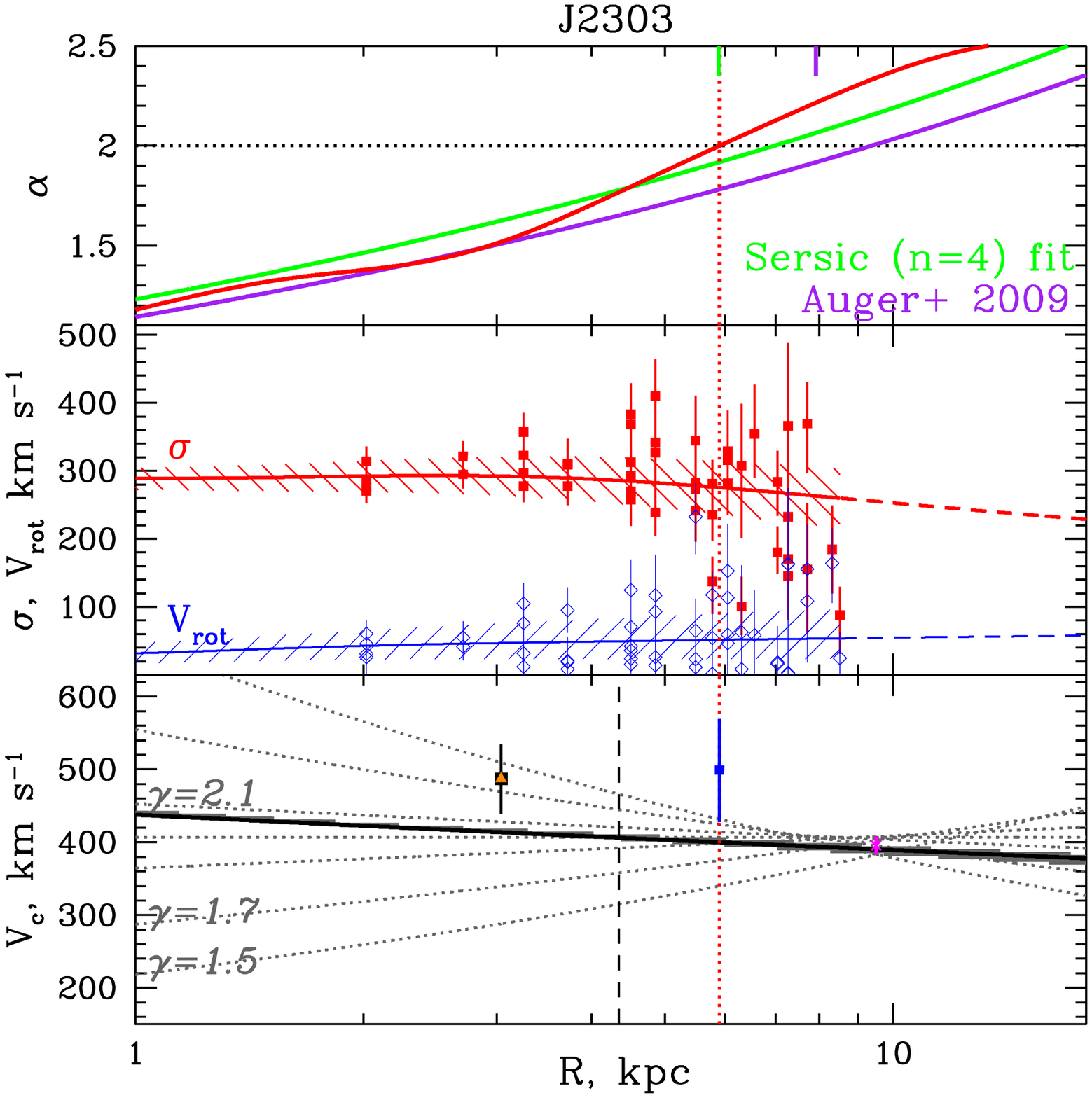}
\plotones{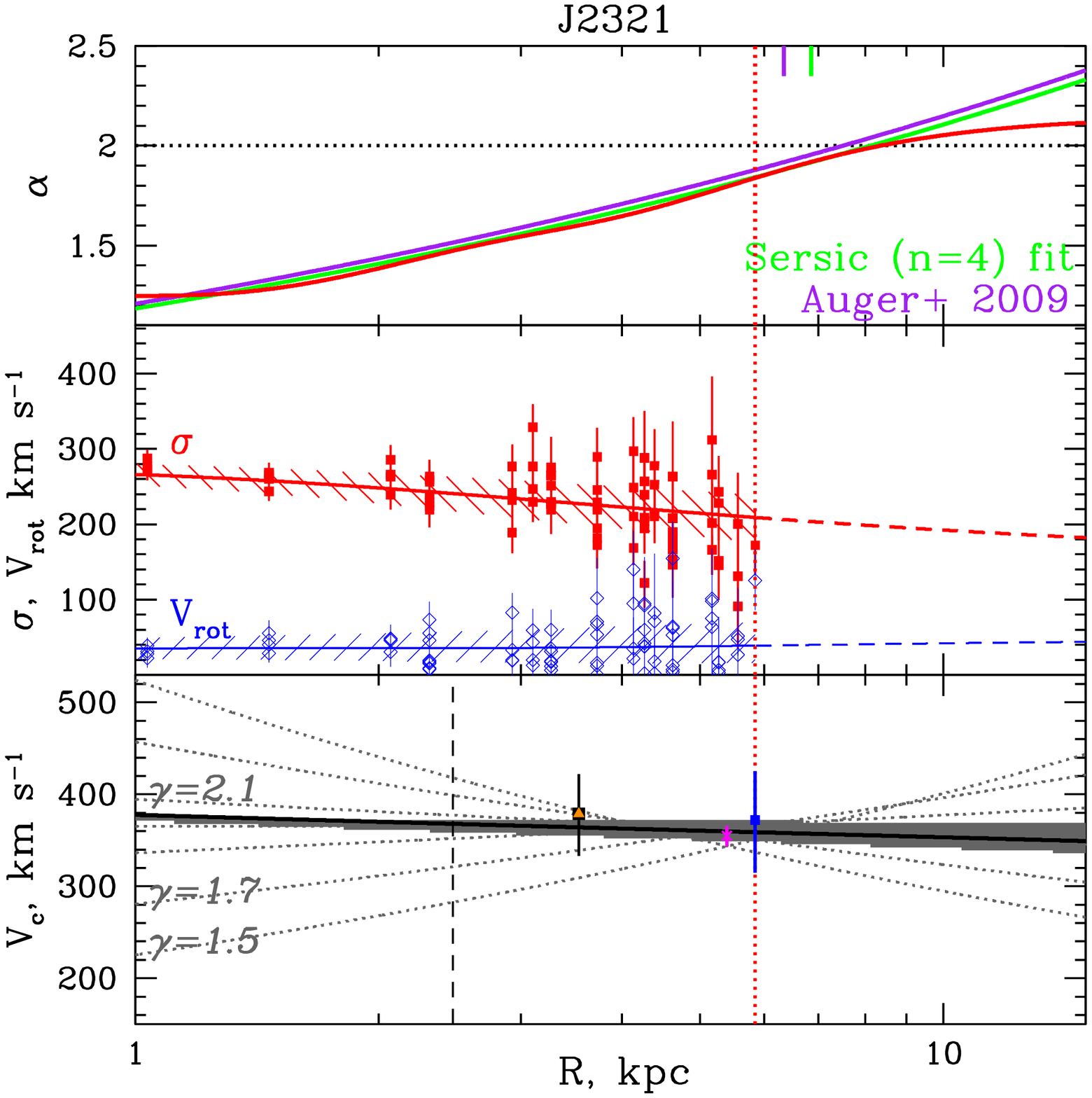}
\caption{(continue)} 
\end{figure*}

\label{lastpage}
\end{document}